\documentclass[article,twocolumn,floatfix,superscriptaddress,longbibliography]{revtex4-2}
\usepackage{physics}
\usepackage[colorlinks,citecolor=blue,linkcolor=blue,urlcolor=blue]{hyperref}
\setcitestyle{super}
\usepackage{graphicx}
\usepackage{color}

\usepackage[colorlinks,citecolor=blue,linkcolor=blue,urlcolor=blue]{hyperref}
\begin{document}

\title{Down-conversion of a single photon as a probe of many-body localization}

\author{Nitish Mehta}
\affiliation{Department of Physics,
University of Maryland, College Park, MD 20742, USA}

\author{Roman Kuzmin}
\affiliation{Department of Physics, University of Maryland, College Park, MD 20742, USA}
\author{ Cristiano Ciuti}
\affiliation{Laboratoire Mat\'{e}riaux et Ph\'{e}nom\`{e}nes Quantiques, Universit\'{e} Paris Cit\'e, CNRS-UMR7162, 75013 Paris, France}

\author{Vladimir E. Manucharyan}
\affiliation{Department of Physics, University of Maryland, College Park, MD 20742, USA}

\begin{abstract}
\noindent
Decay of a particle into more particles is a ubiquitous phenomenon to interacting quantum systems, taking place in colliders, nuclear reactors, or solids. 
In a non-linear medium, even a single photon would decay by down-converting (splitting) into lower-frequency photons with the same total energy~\cite{klyshko1969scattering}, at a rate given by Fermi's Golden Rule~\cite{dirac1927quantum}. However, the energy conservation condition cannot be matched precisely if the medium is finite and only supports quantized modes.  In this case, the photon's fate becomes the long-standing question of many-body localization (MBL), originally formulated as a gedanken experiment for the lifetime of a single Fermi-liquid quasiparticle confined to a quantum dot~\cite{Altschuler1997}. Here we implement such an experiment using a superconducting multi-mode cavity, the non-linearity of which was tailored to strongly violate the photon number conservation. The resulting interaction attempts to convert a single photon excitation into a shower of low-energy photons, but fails due to the MBL mechanism, which manifests as a striking spectral fine structure of multi-particle resonances at the cavity's standing wave mode frequencies. Each resonance was identified as a many-body state of radiation composed of photons from a broad frequency range,and not obeying the Fermi's Golden Rule theory. Our result introduces a new platform to explore fundamentals of MBL without having to control many atoms or qubits~\cite{Schreiber2015,Choi2016, Smith2016, Roushan2017, Xu2019, Lukin2019, bluvstein2021controlling, morong2021observation, guo2021observation}.

\end{abstract}
\date{\today}
\maketitle
MBL was theoretically introduced as an extension of Anderson localization to interacting electrons~\cite{anderson1958absence, gornyi2005interacting, Basko2006, Basko2007}. It was shown that the dynamics of such a complex many-body problem can be mapped onto hopping of a single fictional particle between sites in the Fock space of electronic occupation numbers. Each site represents a many-body eigenstate of non-interacting electrons, while the electron-electron interaction determines the hopping amplitude that couples sites with a sufficiently small energy mismatch but possibly very different particle content. Absence of diffusion in the Fock space was coined MBL, by analogy with Anderson localization in the real space~\cite{Basko2006}, and this general phenomenon was subsequently predicted to occur in many other physical settings~\cite{Oganesyan2007, serbyn2013local, huse2014phenomenology}. Being the only known mechanism to prevent isolated interacting systems from reaching thermal equilibrium, MBL has been attracting a vast attention from the theoretical physics community~\cite{nandkishore_review, Abanin2019}.

Observing MBL in an experiment is a profoundly hard problem. To start, one must isolate from the external environment and individually control a sufficiently large number of interacting quantum degrees of freedom. Most MBL experiments have thus been limited to technologically sophisticated setups, involving either cold atoms \cite{Schreiber2015,Choi2016,Lukin2019, bluvstein2021controlling} in optical lattices or quantum information processors based on trapped ions \cite{Smith2016, morong2021observation} or superconducting qubits \cite{Roushan2017,Xu2019, guo2021observation}. Furthermore, many signatures of the absence of thermalization are qualitatively similar to the behavior of non-interacting systems, such as slow dynamics, suppressed transport, or Poisson energy level statistics \cite{MIRLIN2000259,Serbyn_2016} (lack of level repulsion). Consequently, interpreting the observations in terms of MBL usually requires detailed many-body simulations of the system in question, the difficulty of which grows exponentially with the system size. Therefore, there is a motivation to explore alternative and possibly simpler experimental configurations that can nevertheless exhibit the key features of MBL. 

Curiously, the idea of Anderson localization in the Fock space was first introduced for a conceptually much simpler setup 25 years ago~\cite{Altschuler1997}, long before the term MBL was even coined~\cite{Basko2006}. Namely, the lifetime of a single Fermi-liquid quasiparticle confined to an isolated quantum dot was considered. In the absence of confinement, a quasiparticle of energy $E$ (with respect to the Fermi energy $E_F$) would decay due to electron-electron interaction into a particle-hole shower at a rate $\sim E^2/E_F$, given by Fermi’s Golden Rule. Confinement makes the spectrum of final multi-particle states discrete, which seemingly violates the energy conservation during the decay and questions the general applicability of Fermi’s Golden Rule to isolated interacting systems. A prediction was made that Fermi's Golden Rule theory indeed breaks down and the decay is inhibited in a broad energy range due to Anderson localization in the space of the dot's orbital occupation numbers. The same mechanism was later invoked to make the very first prediction of MBL in extended systems. In theory, Fermi's Golden Rule breakdown would manifest by the splitting of the dot's electronic orbitals into a forest of coherent multi-particle tunneling resonances. Such resonances are now understood to be a generic feature of local spectral functions of MBL systems even in the presence of a moderate coupling to an external  bath~\cite{Huse_PRB_2014, Jori_PRL_2015} and yet they have never been reported experimentally.

\begin{figure*}[t!]
    \centering
    \includegraphics[width=0.8\linewidth]{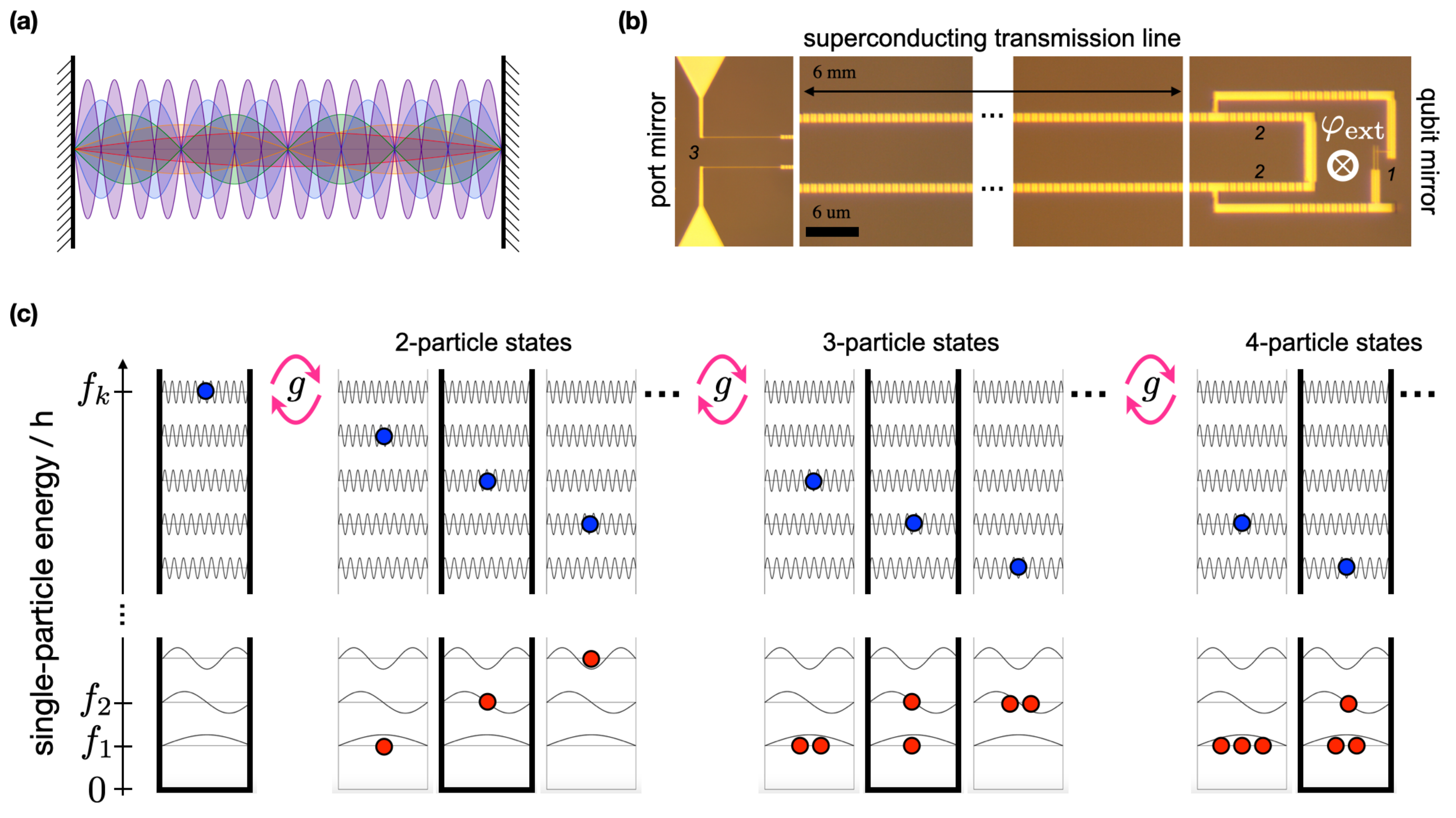}
    \caption{(a) Discrete electromagnetic modes of a long Fabry-Perot cavity. (b) Optical image of the measured device, which consists of a superconducting transmission line terminated on the right side by a fluxonium qubit and on the left side by a dipole antenna (marked `3'). The fluxonium shares half of its inductance with the transmission line (marked `2') and it is tuned by a magnetic flux $\varphi_{\textrm{ext}}\times \hbar/2e$ through the loop. The main junction of fluxonium is marked `1'. (c) A tree of multi-particle states of radiation in the cavity with energies near that of a single photon in mode $k \gg 1$. The qubit induces a photon-photon interaction that coherently hybridizes multi-particle states with the total occupation number differing by a unity.  
    }
    \label{Fig1}
\end{figure*}

We demonstrate the elementary MBL scenario introduced in Ref.~\onlinecite{Altschuler1997} by substituting electrons with photons and the quantum dot with a Fabry-Perot cavity (Fig.~1a). This somewhat uncommon substitution was made possible by our synthesis of an efficient non-linearity in the cavity that can induce a rapid spontaneous down-conversion of a single photon. The cavity (Fig.~1b) consists of a long superconducting transmission line, terminated by a weakly coupled probe antenna at the left end (``port" mirror), and by a superconducting fluxonium qubit~\cite{Manucharyan2009} at the right end (``qubit" mirror). Circuit parameters are chosen such that qubit and cavity interact in the superstrong coupling regime~\cite{Meystre_2006}.
The qubit frequency can be tuned with an externally applied magnetic flux $\varphi_{\textrm{ext}}\times \hbar/2e$ through the circuit loop. The antenna works as a weakly transparent mirror to probe the spectrum of the cavity's standing wave modes. Reflection at the qubit end is solely responsible for the non-linearity.

\begin{figure*}
    \centering
    \includegraphics[width =\linewidth]{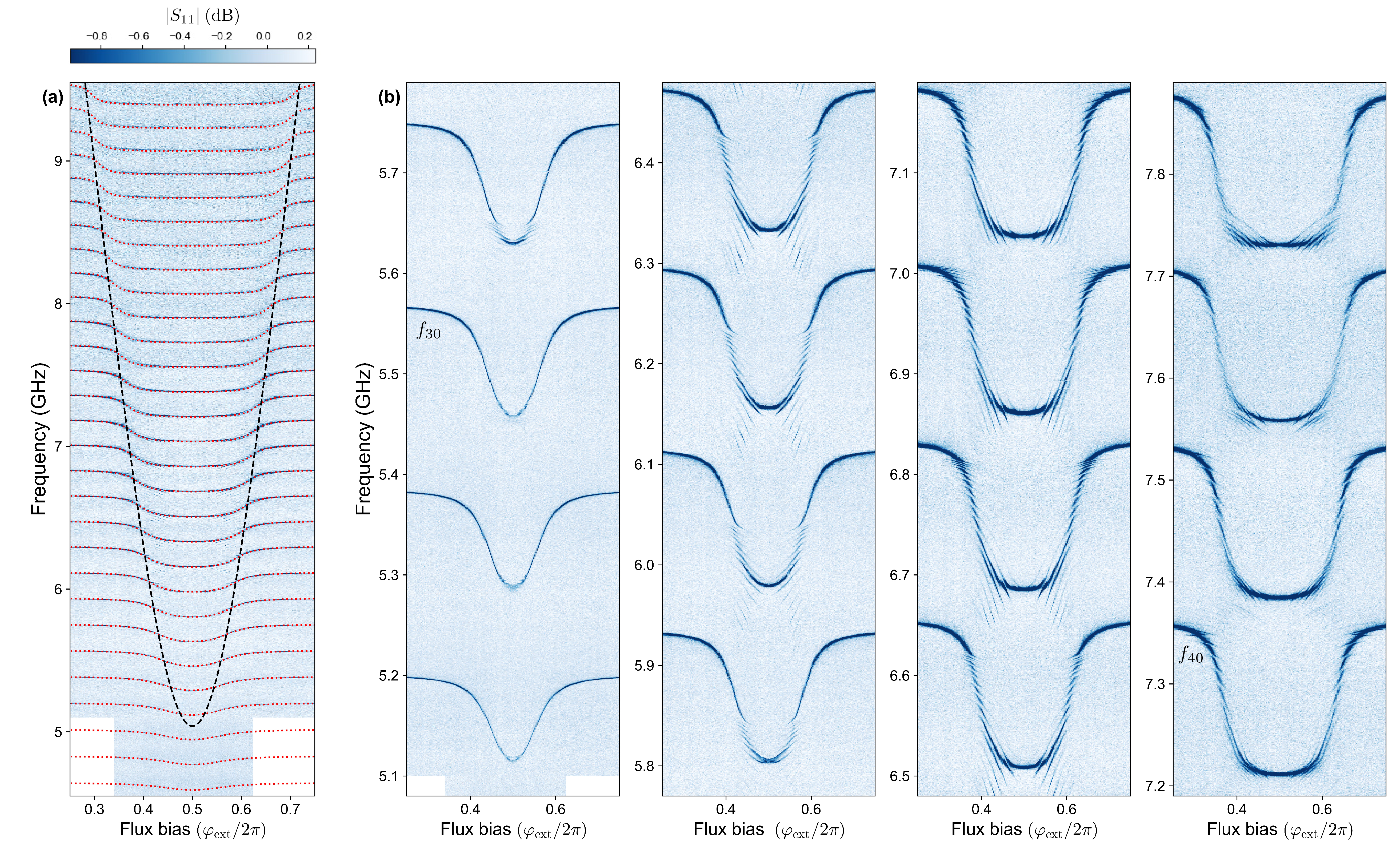}
    \caption{(a) Linear (low-power limit) one-port reflection magnitude $|S_{11}| (\mathrm{dB})$ as a function of excitation frequency and the external flux $\varphi_{\textrm{ext}}$. The red dotted lines depict the fitted cavity spectrum in the single-particle approximation. The black dashed line shows the extracted bare qubit frequency $f_{\textrm{eg}}(\varphi_{\textrm{ext}})$.
    (b) Zoom-in on the cavity modes $k = 28 - 43$ revealing a fine spectral structure around modes with $k > 30$.}
    \label{Fig2}
\end{figure*}

To understand the origin of many-body dynamics in our cavity QED system, let us replace the qubit by an ideal mirror, for the time being, and examine the Fock space of a cavity with uniformly-spaced mode frequencies $f_k = k\Delta$, where $ k=1,2,...$ (Fig.~1a,c). Consider exciting a single high-frequency photon in the mode $k \gg 1$. This single-particle state would be degenerate with about $O(k)$ two-particle states, as well as $O(k^2)$ three-particle states and so on so forth, as illustrated with blue and red color dots in Fig.~1c. What would be the effect of a qubit at frequency $f_{\mathrm{eg}} \sim f_k$? In the single-particle approximation, the qubit merely adds an energy-dependent phase-shift to reflected photons. This linear effect~\cite{Houck_2015,  PuertasMartnez2019, Kuzmin2019} amounts to weakly perturbing the standing wave modes in the frequency window $|f_k - f_{\mathrm{eg}}|\lesssim \Gamma$, where $\Gamma$ coincides with the qubit's spontaneous emission linewidth in the limit of infinite system size and satisfies $\Delta \ll\Gamma \ll f_{\mathrm{eg}}$ (the superstrong coupling condition). At the multi-particle level, though, the qubit also induces a three-wave mixing type interaction described by the following effective Hamiltonian~\cite{Nitish_paper1}:

\begin{equation}
    \frac{\hat{H}}{h} = \sum_{k>0} f_k \hat{a}_k^{\dagger} \hat{a}_k + \hat{V},
    \label{Heff1}
\end{equation}
\begin{equation}
    \hat{V} =g \sum_{j\leq j_0}^ {k,k'>j_0} \sqrt{j} \,A_{k,k'} \hat{a}_j^{\dagger} \hat{a}_{k'}^{\dagger} \hat{a}_k + h.c.\,.
    \label{H_eff}
\end{equation}
Here $\hat{a}_k$ ($\hat{a}_k^{\dagger}$) is the operator that annihilates (creates) a photon at frequency $f_k$ in the standing-wave mode $k$, which \textit{already} includes a small qubit admixture (see Supplementary Material for additional details about the theory). The qubit as a separate entity is completely eliminated from this picture. The 
interaction energy scale $g$ is linked to $\Gamma$ via circuit theory, it vanishes at the fluxonium's flux-inversion symmetry spots, $\varphi_{\textrm{ext}} = 0,\pi$, and satisfies $g \ll \Delta$. The matrix elements $A_{k,k'}$ are of order unity for $k' = k$ but rapidly decay for $|k-k'|\gtrsim \Gamma/\Delta$. The index $j_0$ formally separates low-frequency (red dots in Fig.~1c) and high-frequency (blue dots in Fig.~1c) modes and it can be safely set to $j_0 \sim (f_{\mathrm{eg}}/\Delta)/2$ as long as $\Gamma \ll f_{\mathrm{eg}}$. 

The interaction term $\hat{V}$ in Eq.~(2) coherently couples a single-particle state $\hat{a}_k^{\dagger}|0\rangle$ to two-particle states $\hat{a}_{k'}^{\dagger}\hat{a}_j^{\dagger}|0\rangle$ composed of one high-frequency photon in mode $k'>j_0$ and one low-frequency photon in mode $j\leq j_0$ (Fig.~1c). Likewise, each of these two-particle states can couple to a subset of three-particle states $\hat{a}_{k''}^{\dagger}\hat{a}_{j'}^{\dagger}\hat{a}_j^{\dagger}|0\rangle$, which retain the low-frequency photon in mode $j$ from the previous generation, i.e. the low-frequency photons are stable, and so forth.  Therefore, the relevant many-body Fock space of Hamiltonian (1,2) has a tree-like connectivity, similarly to the case of electron-electron interaction in a quantum dot. An example of a hopping trajectory is highlighted bold in Fig.~1c. If the hopping leads to delocalization along the tree, an initially excited photon in mode $k$ would irreversibly decay by down-conversion at a rate $\propto \sum_{k',j} |\langle 0 \vert \hat{a}_k \hat V \hat{a}_{k'}^{\dagger}\hat{a}_j^{\dagger}\vert 0\rangle|^2\delta(f_k-f_{k'}-f_{j})$ given by Fermi's Golden Rule. The opposite case of localization is the subject of the present experiment.

In addition to interactions, we must consider the effect of disorder in the frequency space, which would lift the multi-particle degeneracy even for $\hat{V}=0$. The natural disorder is negligible in our system, but its role is taken up by dispersion, which comes from three separate mechanisms. In the low-frequency limit, there is a positive dispersion due to the capacitive boundary condition at the antenna end. In the high-frequency limit, the dispersion is negative due to the plasma resonance in the Josephson junctions forming our transmission line. Finally, even in a non-dispersive transmission line, the qubit hybridization increases the mode density around the frequency $f_{\mathrm{eg}}$ by a factor of about  $1+\Delta/\Gamma$. For $k \gg 1$, these dispersive effects altogether quasi-randomly distribute the multi-particle states in energy, thus mimicking a generic disordered system.

Our experiment consists of measuring the reflection magnitude $|S_{11}|$ at the port end of the cavity, following previously established procedures~\cite{Kuzmin2019_1}. The incident power excites the cavity with less than one photon, on average, and we checked that the resulting spectra are power-independent. Let us first discuss the coarse frequency resolution data (Fig.~2a). It shows the expected sequence of standing wave resonances for modes $k=25,26,...,50$, whose frequencies $f_k$ are modulated by the magnetic flux $\varphi_{\textrm{ext}}$ due to the qubit admixture. The flux-modulation rapidly vanishes for modes with $k < 25$. The mode spacing $\Delta$ varies in the range $160-180~\textrm{MHz}$. The measured values $f_k$ vs. $\varphi_{\textrm{ext}}$ agree with a single-particle (linear) hybridization model detailed in Ref.~\cite{Nitish_paper1} (Fig.~2a, dotted lines). Fitting this model provides reliable estimates of all circuit parameters (see the details about the fitting procedure and Table S1 in the Supplemental Material). For example, we can reconstruct the uncoupled qubit frequency (Fig.~2a, dashed line) and estimate $\Gamma \approx 1~\mathrm{GHz}$. The flux knob can continuously modify our system, including switching off the interaction $\hat{V}$ at $\varphi_{\textrm{ext}} = 0,\pi$. In addition to adjusting the single-particle spectrum, we can
maximize the interaction effects for a given mode $k$ by setting $f_{\mathrm{eg}}(\varphi_{\textrm{ext}})\approx f_k \pm \Gamma$. This condition graphically corresponds to the sloped region of $f_k(\varphi_{\textrm{ext}})$ in Fig.~2a.

A higher resolution data reveals dramatic deviations from the single-particle picture (Fig.~2b). Namely, the cavity resonances with $k > 30$ anticross many new flux-dependent lines. The splitting size varies from line to line, but remains much smaller than $\Delta$. The most unusual property of these new spectral lines is that their density is much larger than $1/\Delta$, especially towards higher frequencies. Indeed, where would so many states possibly come from in a cavity excited by at most one photon? In fact, these are precisely the multi-particle states introduced in Fig.~1c. In fact, one can use the $f_k$-data to duly check that each new resonance around the modes $k=31-39$ in Fig.~2b satisfies the two-particle matching condition $f_k = f_{k'}+f_{j}$, for selected integer pairs $k',j < k$. For $k \leq 30$ there are no matching two-particle states in our system irrespective of the flux bias. 

Zooming in to two modes ($k=35, 37$) we find a remarkably good agreement between data and the two-particle spectrum of Hamiltonian (1,2), produced without adjustable parameters (Fig.~3). The matrix elements of $\hat{V}$ were calculated using the circuit parameters obtained from fitting the single-particle model~\cite{Nitish_paper1}.  
The theoretical $|S_{11}|$-signal in Fig.~3 involves the two-particle wavefunctions as well as the values of the intrinsic and extrinsic quality factors of the cavity resonances, which were measured at the non-interacting flux bias $\varphi_{\textrm{ext}}=0$. We stress that linear spectroscopy only lights up states with a substantial single-particle component, which is why the two-particle resonances appear as a ``fine structure" around the cavity mode frequencies. 

The illustrative spectra shown in Fig.~3 are inconsistent with Fermi's Golden Rule decay rate theory, but they have a straightforward interpretation in terms of localization in the many-body Fock space. We focus on a specific realization of our system at $\varphi_{\textrm{ext}}/ 2\pi = 0.356$ (Fig.~3 - insets). Thus, mode $k=35$ appears as a single resonance (as well as all modes $k<35$) and it is accurately described by a single-particle wavefunction, $ \vert \Psi_{\alpha} \rangle \approx \hat{a}_{35}^{\dagger}|0\rangle$. This state is trivially localized in the single-particle sense: a photon excited in mode $k=35$ does not make any attempt to split because the nearest available two-particle states are far detuned compared to the interaction scale. By contrast, a slightly higher frequency mode $k=37$ fragments into at least eight resolved resonances. Each of them is described by a many-body-localized wavefunction, for instance (as marked in Fig.~3 - insets), $\vert \Psi_{\beta} \rangle \approx   0.58 \: \hat{a}^{\dagger}_{37}  \vert 0 \rangle 
+0.102 \: \hat{a}^{\dagger}_{25} \hat{a}^{\dagger}_{12} \vert 0 \rangle +  0.184 \: \hat{a}^{\dagger}_{24} \hat{a}^{\dagger}_{13} \vert 0 \rangle +  0.512 \: \hat{a}^{\dagger}_{23} \hat{a}^{\dagger}_{14} \vert 0 \rangle - 0.53 \: \hat{a}^{\dagger}_{22} \hat{a}^{\dagger}_{15} \vert 0 \rangle - 0.223 \: \hat{a}^{\dagger}_{21} \hat{a}^{\dagger}_{16} \vert 0 \rangle - 0.125 \: \hat{a}^{\dagger}_{20} \hat{a}^{\dagger}_{17} \vert 0 \rangle$. This wavefunction consists of a superposition of the original single-particle state, with a probability $0.58^2 \approx 34\%$, and several nearly-resonant two-particle states. One may say that a photon excited in mode $k=37$ does make splitting attempts, e.g. into a pair of photons in modes $22,15$ or $23, 14$. However, since all the second generation photons correspond to trivially localized states ($k < 30$), the photon shower cascade is blocked in the third generation, and the coherent down-conversion dynamics extends to only a few two-particle sites in the Fock space.  

\begin{figure}[t!]
    \centering
    \includegraphics[width=1\linewidth]{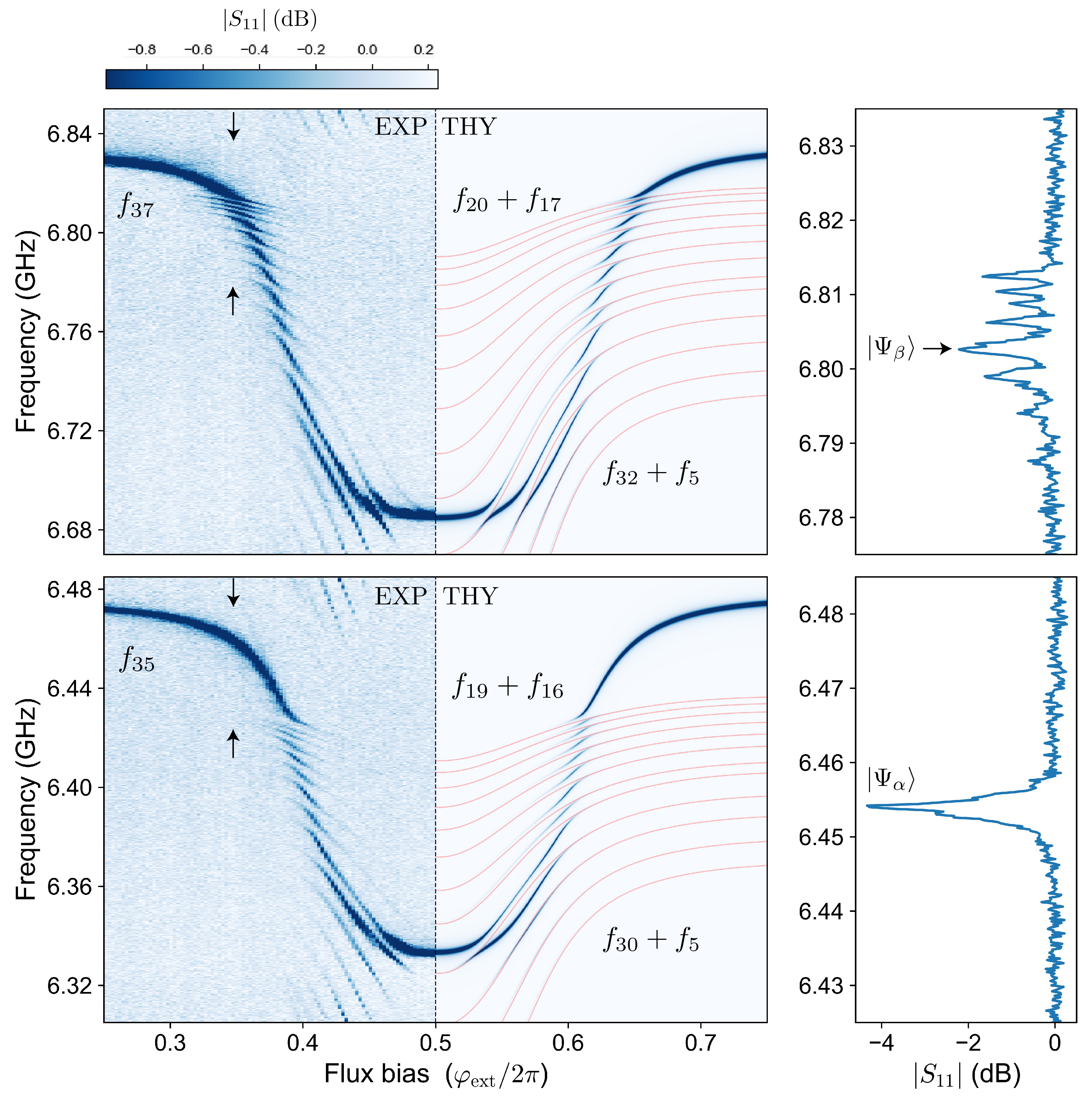}
    \caption{Reflection magnitude $S_{11}$ ($\mathrm{dB}$) near modes $k=35$ and $k=37$, data for $\varphi_{\mathrm{ext}}/ 2 \pi < 0.5$, theory in the two-particle approximation without adjustable parameters for $\varphi_{\mathrm{ext}}/2 \pi > 0.5$. The red lines show uncoupled two-particle energy levels taken directly from the lower-frequency data. The insets show a line cut for an illustrative flux bias $\varphi_{\textrm{ext}}/2 \pi = 0.356$.}
    \label{Fig3}
\end{figure}

\begin{figure*}[t!]
    \centering
    \includegraphics[width
    =0.82\linewidth]{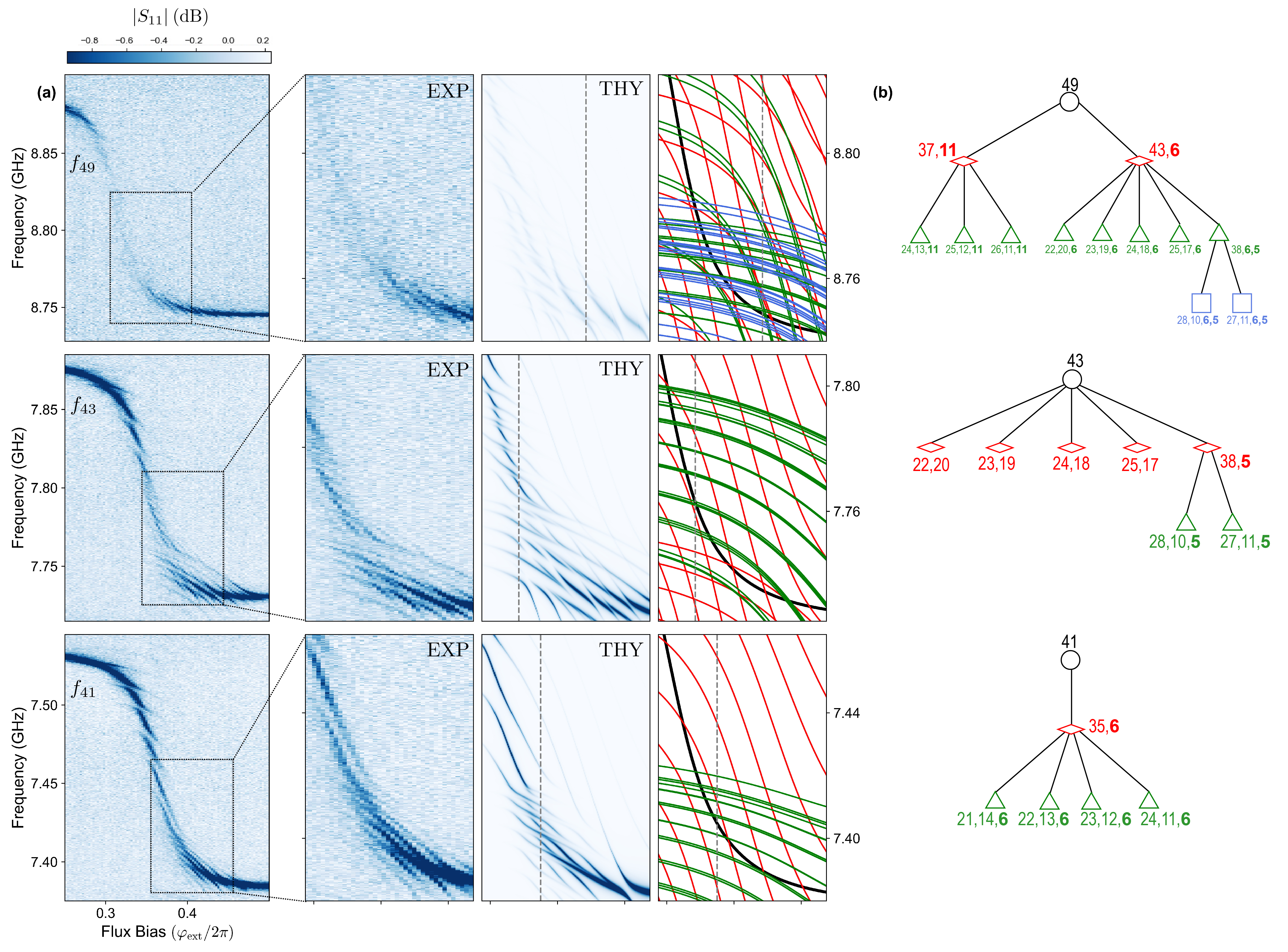}
    \caption{(a) Reflection magnitude $|S_{11}| (\mathrm{dB})$ near modes $49$, $43$ and $41$. Insets (left to right): (left) zoom on the spectroscopy data in the $100~\textrm{MHz}$ wide windows; (middle) calculated spectrum, no adjustable parameters; (right) energies of the uncoupled single-particle (black), two-particle (red), three-particle (green), and four-particle (blue) states used in the calculation.(b) Examples of many-body Fock space trees describing the localization dynamics at selected flux values, indicated by the vertical dashed line in the insets of (a). Note, density of interconnected states is is much smaller than that of all multi-particle states.}
    \label{Fig4}
\end{figure*}

The extent of localization rapidly decreases for higher-frequency modes (Fig.~4a). In particular, the data around modes $k=41,43$ contains extra-fine spectral  features that cannot be explained within the two-particle approximation, but agree with the three-particle spectrum of Hamiltonian (1,2). We illustrate the composition of observed many-body resonances using the Fock-space hopping trees. For clarity, we retain only the strongest branches, for which the hopping amplitude (the modulus of the matrix element of $\hat{V}$) exceeds the energy mismatch. In the example of mode $k=41$ (at $\varphi_{\textrm{ext}} = 0.39$), a single-particle state $\hat{a}_{41}^{\dagger}\vert 0\rangle$ hybridizes predominantly with one two-particle state $\hat{a}_{35}^{\dagger} \hat{a}_{6}^{\dagger}\vert 0\rangle$, which itself hybridizes with four three-particle states 
$\hat{a}_{21}^{\dagger}\hat{a}_{14}^{\dagger}\hat{a}_{6}^{\dagger} \vert 0\rangle$, 
$\hat{a}_{22}^{\dagger}\hat{a}_{13}^{\dagger}\hat{a}_{6}^{\dagger} \vert 0\rangle$, 
$\hat{a}_{23}^{\dagger}\hat{a}_{12}^{\dagger}\hat{a}_{6}^{\dagger} \vert 0\rangle$, 
$\hat{a}_{24}^{\dagger}\hat{a}_{11}^{\dagger}\hat{a}_{6}^{\dagger} \vert 0\rangle$. 
Here the photon cascade is blocked in the fourth generation, because all the third-generation photons ($k\leq 21$) belong to trivially localized states (see Fig.~2b). 
Same analysis applies to another example spectrum around mode $k=43$ ($\varphi_{\textrm{ext}} = 0.367$).

As the density of interconnected multi-particle states grows with energy, one may wonder if Fermi's Golden Rule recovers for higher-frequency modes. We note, though, that our system has an inevitable coupling to an external bath, albeit relatively weak, due to leaking of photons into the measurement apparatus and materials absorption. Therefore, as the hybridization extends to larger and larger number of multi-particle states (hence, lesser degree of many-body localization), their frequency spacing would eventually become comparable to their bath-defined linewidth, which in our system is slightly below $1~\textrm{MHz}$. In this case, the fine spectral structure would merge into a single broad resonance, effectively recovering the applicability of Fermi's Golden Rule in the otherwise localized system. This effect limits our present resolution to two- and three-particle states, as can be seen by the example of mode $k=49$, where we expect the four-particle states to appear in theory (Fig.~4), but instead measure a barely visible broadened resonance.

To summarize, we showed that spontaneous down-conversion of a single photon confined between two mirrors is a many-body phenomenon and it is subject to many-body localization (MBL). Localization leads to a breakdown of Fermi's Golden Rule description of the down-conversion rate, by analogy with the original prediction for a confined Fermi-liquid quasiparticle~\cite{Altschuler1997}. The cavity's standing wave modes fragment into a spectacular fine structure of multi-particle resonances, which unambiguously distinguishes MBL from either the delocalization regime (single interaction-broadened resonance) or the non-interacting photons (single sharp resonance).

Our demonstration puts forward multi-mode circuit quantum electrodynamics
as a possibly useful experimental platform to further explore fundamentals of MBL. In comparison to spin chains or lattice fermions, here the frequency space takes the role of the real space, the number of participating degrees of freedom is set by the excitation frequency, and the connectivity of the many-body Fock space can be tailored by varying the properties of the qubit circuit, serving as one of the cavity mirrors. In the present experiment, we implemented a local tree-like connectivity by harnessing the inversion symmetry breaking and the strong anharmonicity properties of fluxoniums. By contrast, the weak anharmonicity of transmons would preserve the particle number and hence 
produces negligible interactions in the many-body sector \cite{Nigg_2012,Kuzmin2019}.  At the opposite extreme is a quantum phase-slip center, which can maximize the Fock-space connectivity, in which case the down-conversion rate indeed accurately obeys the Fermi's Golden Rule \cite{Kuzmin_2021}. In future experiments, the disorder can be tailored via nanofabrication, the interaction can be quenched by rapidly tuning the flux knob, 
and multi-mode tomography techniques, developed for bosonic qubits \cite{gao2019entanglement}, can be applied here to monitor the entanglement dynamics in the time-domain. 

\acknowledgements
{We acknowledge support from DOE Early Career Award and from the ARO-MURI program. C.C. acknowledges financial support from FET FLAGSHIP Project PhoQuS (grant agreement ID no.820392) and from the French agency ANR through the project NOMOS (ANR-18-CE24-0026), and TRIANGLE (ANR-20-CE47-0011).}
	
\bibliography{biblio}

%apsrev4-2.bst 2019-01-14 (MD) hand-edited version of apsrev4-1.bst
%Control: key (0)
%Control: author (8) initials jnrlst
%Control: editor formatted (1) identically to author
%Control: production of article title (0) allowed
%Control: page (0) single
%Control: year (1) truncated
%Control: production of eprint (0) enabled
\begin{thebibliography}{35}%
\makeatletter
\providecommand \@ifxundefined [1]{%
 \@ifx{#1\undefined}
}%
\providecommand \@ifnum [1]{%
 \ifnum #1\expandafter \@firstoftwo
 \else \expandafter \@secondoftwo
 \fi
}%
\providecommand \@ifx [1]{%
 \ifx #1\expandafter \@firstoftwo
 \else \expandafter \@secondoftwo
 \fi
}%
\providecommand \natexlab [1]{#1}%
\providecommand \enquote  [1]{``#1''}%
\providecommand \bibnamefont  [1]{#1}%
\providecommand \bibfnamefont [1]{#1}%
\providecommand \citenamefont [1]{#1}%
\providecommand \href@noop [0]{\@secondoftwo}%
\providecommand \href [0]{\begingroup \@sanitize@url \@href}%
\providecommand \@href[1]{\@@startlink{#1}\@@href}%
\providecommand \@@href[1]{\endgroup#1\@@endlink}%
\providecommand \@sanitize@url [0]{\catcode `\\12\catcode `\$12\catcode
  `\&12\catcode `\#12\catcode `\^12\catcode `\_12\catcode `\%12\relax}%
\providecommand \@@startlink[1]{}%
\providecommand \@@endlink[0]{}%
\providecommand \url  [0]{\begingroup\@sanitize@url \@url }%
\providecommand \@url [1]{\endgroup\@href {#1}{\urlprefix }}%
\providecommand \urlprefix  [0]{URL }%
\providecommand \Eprint [0]{\href }%
\providecommand \doibase [0]{https://doi.org/}%
\providecommand \selectlanguage [0]{\@gobble}%
\providecommand \bibinfo  [0]{\@secondoftwo}%
\providecommand \bibfield  [0]{\@secondoftwo}%
\providecommand \translation [1]{[#1]}%
\providecommand \BibitemOpen [0]{}%
\providecommand \bibitemStop [0]{}%
\providecommand \bibitemNoStop [0]{.\EOS\space}%
\providecommand \EOS [0]{\spacefactor3000\relax}%
\providecommand \BibitemShut  [1]{\csname bibitem#1\endcsname}%
\let\auto@bib@innerbib\@empty
%</preamble>
\bibitem [{\citenamefont {Klyshko}(1969)}]{klyshko1969scattering}%
  \BibitemOpen
  \bibfield  {author} {\bibinfo {author} {\bibfnamefont {D.}~\bibnamefont
  {Klyshko}},\ }\bibfield  {title} {\bibinfo {title} {Scattering of light in a
  medium with nonlinear polarizability},\ }\href
  {http://www.jetp.ras.ru/cgi-bin/dn/e_028_03_0522.pdf} {\bibfield  {journal}
  {\bibinfo  {journal} {Sov. Phys. JETP}\ }\textbf {\bibinfo {volume} {28}},\
  \bibinfo {pages} {522} (\bibinfo {year} {1969})}\BibitemShut {NoStop}%
\bibitem [{\citenamefont {Dirac}(1927)}]{dirac1927quantum}%
  \BibitemOpen
  \bibfield  {author} {\bibinfo {author} {\bibfnamefont {P.~A.~M.}\
  \bibnamefont {Dirac}},\ }\bibfield  {title} {\bibinfo {title} {The quantum
  theory of the emission and absorption of radiation},\ }\href
  {http://doi.org/10.1098/rspa.1927.0039} {\bibfield  {journal} {\bibinfo
  {journal} {Proceedings of the Royal Society of London. Series A, Containing
  Papers of a Mathematical and Physical Character}\ }\textbf {\bibinfo {volume}
  {114}},\ \bibinfo {pages} {243} (\bibinfo {year} {1927})}\BibitemShut
  {NoStop}%
\bibitem [{\citenamefont {Altshuler}\ \emph {et~al.}(1997)\citenamefont
  {Altshuler}, \citenamefont {Gefen}, \citenamefont {Kamenev},\ and\
  \citenamefont {Levitov}}]{Altschuler1997}%
  \BibitemOpen
  \bibfield  {author} {\bibinfo {author} {\bibfnamefont {B.~L.}\ \bibnamefont
  {Altshuler}}, \bibinfo {author} {\bibfnamefont {Y.}~\bibnamefont {Gefen}},
  \bibinfo {author} {\bibfnamefont {A.}~\bibnamefont {Kamenev}},\ and\ \bibinfo
  {author} {\bibfnamefont {L.~S.}\ \bibnamefont {Levitov}},\ }\bibfield
  {title} {\bibinfo {title} {Quasiparticle lifetime in a finite system: A
  nonperturbative approach},\ }\href
  {https://doi.org/10.1103/PhysRevLett.78.2803} {\bibfield  {journal} {\bibinfo
   {journal} {Phys. Rev. Lett.}\ }\textbf {\bibinfo {volume} {78}},\ \bibinfo
  {pages} {2803} (\bibinfo {year} {1997})}\BibitemShut {NoStop}%
\bibitem [{\citenamefont {Schreiber}\ \emph {et~al.}(2015)\citenamefont
  {Schreiber}, \citenamefont {Hodgman}, \citenamefont {Bordia}, \citenamefont
  {L\"{u}schen}, \citenamefont {Fischer}, \citenamefont {Vosk}, \citenamefont
  {Altman}, \citenamefont {Schneider},\ and\ \citenamefont
  {Bloch}}]{Schreiber2015}%
  \BibitemOpen
  \bibfield  {author} {\bibinfo {author} {\bibfnamefont {M.}~\bibnamefont
  {Schreiber}}, \bibinfo {author} {\bibfnamefont {S.~S.}\ \bibnamefont
  {Hodgman}}, \bibinfo {author} {\bibfnamefont {P.}~\bibnamefont {Bordia}},
  \bibinfo {author} {\bibfnamefont {H.~P.}\ \bibnamefont {L\"{u}schen}},
  \bibinfo {author} {\bibfnamefont {M.~H.}\ \bibnamefont {Fischer}}, \bibinfo
  {author} {\bibfnamefont {R.}~\bibnamefont {Vosk}}, \bibinfo {author}
  {\bibfnamefont {E.}~\bibnamefont {Altman}}, \bibinfo {author} {\bibfnamefont
  {U.}~\bibnamefont {Schneider}},\ and\ \bibinfo {author} {\bibfnamefont
  {I.}~\bibnamefont {Bloch}},\ }\bibfield  {title} {\bibinfo {title}
  {Observation of many-body localization of interacting fermions in a
  quasirandom optical lattice},\ }\href
  {https://doi.org/10.1126/science.aaa7432} {\bibfield  {journal} {\bibinfo
  {journal} {Science}\ }\textbf {\bibinfo {volume} {349}},\ \bibinfo {pages}
  {842} (\bibinfo {year} {2015})}\BibitemShut {NoStop}%
\bibitem [{\citenamefont {yoon Choi}\ \emph {et~al.}(2016)\citenamefont {yoon
  Choi}, \citenamefont {Hild}, \citenamefont {Zeiher}, \citenamefont
  {Schau{\ss}}, \citenamefont {Rubio-Abadal}, \citenamefont {Yefsah},
  \citenamefont {Khemani}, \citenamefont {Huse}, \citenamefont {Bloch},\ and\
  \citenamefont {Gross}}]{Choi2016}%
  \BibitemOpen
  \bibfield  {author} {\bibinfo {author} {\bibfnamefont {J.}~\bibnamefont {yoon
  Choi}}, \bibinfo {author} {\bibfnamefont {S.}~\bibnamefont {Hild}}, \bibinfo
  {author} {\bibfnamefont {J.}~\bibnamefont {Zeiher}}, \bibinfo {author}
  {\bibfnamefont {P.}~\bibnamefont {Schau{\ss}}}, \bibinfo {author}
  {\bibfnamefont {A.}~\bibnamefont {Rubio-Abadal}}, \bibinfo {author}
  {\bibfnamefont {T.}~\bibnamefont {Yefsah}}, \bibinfo {author} {\bibfnamefont
  {V.}~\bibnamefont {Khemani}}, \bibinfo {author} {\bibfnamefont {D.~A.}\
  \bibnamefont {Huse}}, \bibinfo {author} {\bibfnamefont {I.}~\bibnamefont
  {Bloch}},\ and\ \bibinfo {author} {\bibfnamefont {C.}~\bibnamefont {Gross}},\
  }\bibfield  {title} {\bibinfo {title} {Exploring the many-body localization
  transition in two dimensions},\ }\href
  {https://doi.org/10.1126/science.aaf8834} {\bibfield  {journal} {\bibinfo
  {journal} {Science}\ }\textbf {\bibinfo {volume} {352}},\ \bibinfo {pages}
  {1547} (\bibinfo {year} {2016})}\BibitemShut {NoStop}%
\bibitem [{\citenamefont {Smith}\ \emph {et~al.}(2016)\citenamefont {Smith},
  \citenamefont {Lee}, \citenamefont {Richerme}, \citenamefont {Neyenhuis},
  \citenamefont {Hess}, \citenamefont {Hauke}, \citenamefont {Heyl},
  \citenamefont {Huse},\ and\ \citenamefont {Monroe}}]{Smith2016}%
  \BibitemOpen
  \bibfield  {author} {\bibinfo {author} {\bibfnamefont {J.}~\bibnamefont
  {Smith}}, \bibinfo {author} {\bibfnamefont {A.}~\bibnamefont {Lee}}, \bibinfo
  {author} {\bibfnamefont {P.}~\bibnamefont {Richerme}}, \bibinfo {author}
  {\bibfnamefont {B.}~\bibnamefont {Neyenhuis}}, \bibinfo {author}
  {\bibfnamefont {P.~W.}\ \bibnamefont {Hess}}, \bibinfo {author}
  {\bibfnamefont {P.}~\bibnamefont {Hauke}}, \bibinfo {author} {\bibfnamefont
  {M.}~\bibnamefont {Heyl}}, \bibinfo {author} {\bibfnamefont {D.~A.}\
  \bibnamefont {Huse}},\ and\ \bibinfo {author} {\bibfnamefont
  {C.}~\bibnamefont {Monroe}},\ }\bibfield  {title} {\bibinfo {title}
  {Many-body localization in a quantum simulator with programmable random
  disorder},\ }\href {https://doi.org/10.1038/nphys3783} {\bibfield  {journal}
  {\bibinfo  {journal} {Nature Physics}\ }\textbf {\bibinfo {volume} {12}},\
  \bibinfo {pages} {907} (\bibinfo {year} {2016})}\BibitemShut {NoStop}%
\bibitem [{\citenamefont {Roushan}\ \emph {et~al.}(2017)\citenamefont
  {Roushan}, \citenamefont {Neill}, \citenamefont {Tangpanitanon},
  \citenamefont {Bastidas}, \citenamefont {Megrant}, \citenamefont {Barends},
  \citenamefont {Chen}, \citenamefont {Chen}, \citenamefont {Chiaro},
  \citenamefont {Dunsworth}, \citenamefont {Fowler}, \citenamefont {Foxen},
  \citenamefont {Giustina}, \citenamefont {Jeffrey}, \citenamefont {Kelly},
  \citenamefont {Lucero}, \citenamefont {Mutus}, \citenamefont {Neeley},
  \citenamefont {Quintana}, \citenamefont {Sank}, \citenamefont {Vainsencher},
  \citenamefont {Wenner}, \citenamefont {White}, \citenamefont {Neven},
  \citenamefont {Angelakis},\ and\ \citenamefont {Martinis}}]{Roushan2017}%
  \BibitemOpen
  \bibfield  {author} {\bibinfo {author} {\bibfnamefont {P.}~\bibnamefont
  {Roushan}}, \bibinfo {author} {\bibfnamefont {C.}~\bibnamefont {Neill}},
  \bibinfo {author} {\bibfnamefont {J.}~\bibnamefont {Tangpanitanon}}, \bibinfo
  {author} {\bibfnamefont {V.~M.}\ \bibnamefont {Bastidas}}, \bibinfo {author}
  {\bibfnamefont {A.}~\bibnamefont {Megrant}}, \bibinfo {author} {\bibfnamefont
  {R.}~\bibnamefont {Barends}}, \bibinfo {author} {\bibfnamefont
  {Y.}~\bibnamefont {Chen}}, \bibinfo {author} {\bibfnamefont {Z.}~\bibnamefont
  {Chen}}, \bibinfo {author} {\bibfnamefont {B.}~\bibnamefont {Chiaro}},
  \bibinfo {author} {\bibfnamefont {A.}~\bibnamefont {Dunsworth}}, \bibinfo
  {author} {\bibfnamefont {A.}~\bibnamefont {Fowler}}, \bibinfo {author}
  {\bibfnamefont {B.}~\bibnamefont {Foxen}}, \bibinfo {author} {\bibfnamefont
  {M.}~\bibnamefont {Giustina}}, \bibinfo {author} {\bibfnamefont
  {E.}~\bibnamefont {Jeffrey}}, \bibinfo {author} {\bibfnamefont
  {J.}~\bibnamefont {Kelly}}, \bibinfo {author} {\bibfnamefont
  {E.}~\bibnamefont {Lucero}}, \bibinfo {author} {\bibfnamefont
  {J.}~\bibnamefont {Mutus}}, \bibinfo {author} {\bibfnamefont
  {M.}~\bibnamefont {Neeley}}, \bibinfo {author} {\bibfnamefont
  {C.}~\bibnamefont {Quintana}}, \bibinfo {author} {\bibfnamefont
  {D.}~\bibnamefont {Sank}}, \bibinfo {author} {\bibfnamefont {A.}~\bibnamefont
  {Vainsencher}}, \bibinfo {author} {\bibfnamefont {J.}~\bibnamefont {Wenner}},
  \bibinfo {author} {\bibfnamefont {T.}~\bibnamefont {White}}, \bibinfo
  {author} {\bibfnamefont {H.}~\bibnamefont {Neven}}, \bibinfo {author}
  {\bibfnamefont {D.~G.}\ \bibnamefont {Angelakis}},\ and\ \bibinfo {author}
  {\bibfnamefont {J.}~\bibnamefont {Martinis}},\ }\bibfield  {title} {\bibinfo
  {title} {Spectroscopic signatures of localization with interacting photons in
  superconducting qubits},\ }\href {https://doi.org/10.1126/science.aao1401}
  {\bibfield  {journal} {\bibinfo  {journal} {Science}\ }\textbf {\bibinfo
  {volume} {358}},\ \bibinfo {pages} {1175} (\bibinfo {year}
  {2017})}\BibitemShut {NoStop}%
\bibitem [{\citenamefont {Xu}\ \emph {et~al.}(2018)\citenamefont {Xu},
  \citenamefont {Chen}, \citenamefont {Zeng}, \citenamefont {Zhang},
  \citenamefont {Song}, \citenamefont {Liu}, \citenamefont {Guo}, \citenamefont
  {Zhang}, \citenamefont {Xu}, \citenamefont {Deng}, \citenamefont {Huang},
  \citenamefont {Wang}, \citenamefont {Zhu}, \citenamefont {Zheng},\ and\
  \citenamefont {Fan}}]{Xu2019}%
  \BibitemOpen
  \bibfield  {author} {\bibinfo {author} {\bibfnamefont {K.}~\bibnamefont
  {Xu}}, \bibinfo {author} {\bibfnamefont {J.-J.}\ \bibnamefont {Chen}},
  \bibinfo {author} {\bibfnamefont {Y.}~\bibnamefont {Zeng}}, \bibinfo {author}
  {\bibfnamefont {Y.-R.}\ \bibnamefont {Zhang}}, \bibinfo {author}
  {\bibfnamefont {C.}~\bibnamefont {Song}}, \bibinfo {author} {\bibfnamefont
  {W.}~\bibnamefont {Liu}}, \bibinfo {author} {\bibfnamefont {Q.}~\bibnamefont
  {Guo}}, \bibinfo {author} {\bibfnamefont {P.}~\bibnamefont {Zhang}}, \bibinfo
  {author} {\bibfnamefont {D.}~\bibnamefont {Xu}}, \bibinfo {author}
  {\bibfnamefont {H.}~\bibnamefont {Deng}}, \bibinfo {author} {\bibfnamefont
  {K.}~\bibnamefont {Huang}}, \bibinfo {author} {\bibfnamefont
  {H.}~\bibnamefont {Wang}}, \bibinfo {author} {\bibfnamefont {X.}~\bibnamefont
  {Zhu}}, \bibinfo {author} {\bibfnamefont {D.}~\bibnamefont {Zheng}},\ and\
  \bibinfo {author} {\bibfnamefont {H.}~\bibnamefont {Fan}},\ }\bibfield
  {title} {\bibinfo {title} {Emulating many-body localization with a
  superconducting quantum processor},\ }\href
  {https://doi.org/10.1103/PhysRevLett.120.050507} {\bibfield  {journal}
  {\bibinfo  {journal} {Phys. Rev. Lett.}\ }\textbf {\bibinfo {volume} {120}},\
  \bibinfo {pages} {050507} (\bibinfo {year} {2018})}\BibitemShut {NoStop}%
\bibitem [{\citenamefont {Lukin}\ \emph {et~al.}(2019)\citenamefont {Lukin},
  \citenamefont {Rispoli}, \citenamefont {Schittko}, \citenamefont {Tai},
  \citenamefont {Kaufman}, \citenamefont {Choi}, \citenamefont {Khemani},
  \citenamefont {L{\'{e}}onard},\ and\ \citenamefont {Greiner}}]{Lukin2019}%
  \BibitemOpen
  \bibfield  {author} {\bibinfo {author} {\bibfnamefont {A.}~\bibnamefont
  {Lukin}}, \bibinfo {author} {\bibfnamefont {M.}~\bibnamefont {Rispoli}},
  \bibinfo {author} {\bibfnamefont {R.}~\bibnamefont {Schittko}}, \bibinfo
  {author} {\bibfnamefont {M.~E.}\ \bibnamefont {Tai}}, \bibinfo {author}
  {\bibfnamefont {A.~M.}\ \bibnamefont {Kaufman}}, \bibinfo {author}
  {\bibfnamefont {S.}~\bibnamefont {Choi}}, \bibinfo {author} {\bibfnamefont
  {V.}~\bibnamefont {Khemani}}, \bibinfo {author} {\bibfnamefont
  {J.}~\bibnamefont {L{\'{e}}onard}},\ and\ \bibinfo {author} {\bibfnamefont
  {M.}~\bibnamefont {Greiner}},\ }\bibfield  {title} {\bibinfo {title} {Probing
  entanglement in a many-body{\textendash}localized system},\ }\href
  {https://doi.org/10.1126/science.aau0818} {\bibfield  {journal} {\bibinfo
  {journal} {Science}\ }\textbf {\bibinfo {volume} {364}},\ \bibinfo {pages}
  {256} (\bibinfo {year} {2019})}\BibitemShut {NoStop}%
\bibitem [{\citenamefont {Bluvstein}\ \emph {et~al.}(2021)\citenamefont
  {Bluvstein}, \citenamefont {Omran}, \citenamefont {Levine}, \citenamefont
  {Keesling}, \citenamefont {Semeghini}, \citenamefont {Ebadi}, \citenamefont
  {Wang}, \citenamefont {Michailidis}, \citenamefont {Maskara}, \citenamefont
  {Ho}, \citenamefont {Choi}, \citenamefont {Serbyn}, \citenamefont {Greiner},
  \citenamefont {Vuletić},\ and\ \citenamefont
  {Lukin}}]{bluvstein2021controlling}%
  \BibitemOpen
  \bibfield  {author} {\bibinfo {author} {\bibfnamefont {D.}~\bibnamefont
  {Bluvstein}}, \bibinfo {author} {\bibfnamefont {A.}~\bibnamefont {Omran}},
  \bibinfo {author} {\bibfnamefont {H.}~\bibnamefont {Levine}}, \bibinfo
  {author} {\bibfnamefont {A.}~\bibnamefont {Keesling}}, \bibinfo {author}
  {\bibfnamefont {G.}~\bibnamefont {Semeghini}}, \bibinfo {author}
  {\bibfnamefont {S.}~\bibnamefont {Ebadi}}, \bibinfo {author} {\bibfnamefont
  {T.~T.}\ \bibnamefont {Wang}}, \bibinfo {author} {\bibfnamefont {A.~A.}\
  \bibnamefont {Michailidis}}, \bibinfo {author} {\bibfnamefont
  {N.}~\bibnamefont {Maskara}}, \bibinfo {author} {\bibfnamefont {W.~W.}\
  \bibnamefont {Ho}}, \bibinfo {author} {\bibfnamefont {S.}~\bibnamefont
  {Choi}}, \bibinfo {author} {\bibfnamefont {M.}~\bibnamefont {Serbyn}},
  \bibinfo {author} {\bibfnamefont {M.}~\bibnamefont {Greiner}}, \bibinfo
  {author} {\bibfnamefont {V.}~\bibnamefont {Vuletić}},\ and\ \bibinfo
  {author} {\bibfnamefont {M.~D.}\ \bibnamefont {Lukin}},\ }\bibfield  {title}
  {\bibinfo {title} {Controlling quantum many-body dynamics in driven {R}ydberg
  atom arrays},\ }\href {https://doi.org/10.1126/science.abg2530} {\bibfield
  {journal} {\bibinfo  {journal} {Science}\ }\textbf {\bibinfo {volume}
  {371}},\ \bibinfo {pages} {1355} (\bibinfo {year} {2021})}\BibitemShut
  {NoStop}%
\bibitem [{\citenamefont {Morong}\ \emph {et~al.}(2021)\citenamefont {Morong},
  \citenamefont {Liu}, \citenamefont {Becker}, \citenamefont {Collins},
  \citenamefont {Feng}, \citenamefont {Kyprianidis}, \citenamefont {Pagano},
  \citenamefont {You}, \citenamefont {Gorshkov},\ and\ \citenamefont
  {Monroe}}]{morong2021observation}%
  \BibitemOpen
  \bibfield  {author} {\bibinfo {author} {\bibfnamefont {W.}~\bibnamefont
  {Morong}}, \bibinfo {author} {\bibfnamefont {F.}~\bibnamefont {Liu}},
  \bibinfo {author} {\bibfnamefont {P.}~\bibnamefont {Becker}}, \bibinfo
  {author} {\bibfnamefont {K.}~\bibnamefont {Collins}}, \bibinfo {author}
  {\bibfnamefont {L.}~\bibnamefont {Feng}}, \bibinfo {author} {\bibfnamefont
  {A.}~\bibnamefont {Kyprianidis}}, \bibinfo {author} {\bibfnamefont
  {G.}~\bibnamefont {Pagano}}, \bibinfo {author} {\bibfnamefont
  {T.}~\bibnamefont {You}}, \bibinfo {author} {\bibfnamefont {A.}~\bibnamefont
  {Gorshkov}},\ and\ \bibinfo {author} {\bibfnamefont {C.}~\bibnamefont
  {Monroe}},\ }\bibfield  {title} {\bibinfo {title} {Observation of stark
  many-body localization without disorder},\ }\href
  {https://doi.org/10.1038/s41586-021-03988-0} {\bibfield  {journal} {\bibinfo
  {journal} {Nature}\ }\textbf {\bibinfo {volume} {599}},\ \bibinfo {pages}
  {393} (\bibinfo {year} {2021})}\BibitemShut {NoStop}%
\bibitem [{\citenamefont {Guo}\ \emph {et~al.}(2021)\citenamefont {Guo},
  \citenamefont {Cheng}, \citenamefont {Sun}, \citenamefont {Song},
  \citenamefont {Li}, \citenamefont {Wang}, \citenamefont {Ren}, \citenamefont
  {Dong}, \citenamefont {Zheng}, \citenamefont {Zhang} \emph
  {et~al.}}]{guo2021observation}%
  \BibitemOpen
  \bibfield  {author} {\bibinfo {author} {\bibfnamefont {Q.}~\bibnamefont
  {Guo}}, \bibinfo {author} {\bibfnamefont {C.}~\bibnamefont {Cheng}}, \bibinfo
  {author} {\bibfnamefont {Z.-H.}\ \bibnamefont {Sun}}, \bibinfo {author}
  {\bibfnamefont {Z.}~\bibnamefont {Song}}, \bibinfo {author} {\bibfnamefont
  {H.}~\bibnamefont {Li}}, \bibinfo {author} {\bibfnamefont {Z.}~\bibnamefont
  {Wang}}, \bibinfo {author} {\bibfnamefont {W.}~\bibnamefont {Ren}}, \bibinfo
  {author} {\bibfnamefont {H.}~\bibnamefont {Dong}}, \bibinfo {author}
  {\bibfnamefont {D.}~\bibnamefont {Zheng}}, \bibinfo {author} {\bibfnamefont
  {Y.-R.}\ \bibnamefont {Zhang}}, \emph {et~al.},\ }\bibfield  {title}
  {\bibinfo {title} {Observation of energy-resolved many-body localization},\
  }\href {https://www.nature.com/articles/s41567-020-1035-1} {\bibfield
  {journal} {\bibinfo  {journal} {Nature Physics}\ }\textbf {\bibinfo {volume}
  {17}},\ \bibinfo {pages} {234} (\bibinfo {year} {2021})}\BibitemShut
  {NoStop}%
\bibitem [{\citenamefont {Anderson}(1958)}]{anderson1958absence}%
  \BibitemOpen
  \bibfield  {author} {\bibinfo {author} {\bibfnamefont {P.~W.}\ \bibnamefont
  {Anderson}},\ }\bibfield  {title} {\bibinfo {title} {Absence of diffusion in
  certain random lattices},\ }\href {https://doi.org/10.1103/PhysRev.109.1492}
  {\bibfield  {journal} {\bibinfo  {journal} {Phys. Rev.}\ }\textbf {\bibinfo
  {volume} {109}},\ \bibinfo {pages} {1492} (\bibinfo {year}
  {1958})}\BibitemShut {NoStop}%
\bibitem [{\citenamefont {Gornyi}\ \emph {et~al.}(2005)\citenamefont {Gornyi},
  \citenamefont {Mirlin},\ and\ \citenamefont
  {Polyakov}}]{gornyi2005interacting}%
  \BibitemOpen
  \bibfield  {author} {\bibinfo {author} {\bibfnamefont {I.~V.}\ \bibnamefont
  {Gornyi}}, \bibinfo {author} {\bibfnamefont {A.~D.}\ \bibnamefont {Mirlin}},\
  and\ \bibinfo {author} {\bibfnamefont {D.~G.}\ \bibnamefont {Polyakov}},\
  }\bibfield  {title} {\bibinfo {title} {Interacting electrons in disordered
  wires: Anderson localization and low-$t$ transport},\ }\href
  {https://doi.org/10.1103/PhysRevLett.95.206603} {\bibfield  {journal}
  {\bibinfo  {journal} {Phys. Rev. Lett.}\ }\textbf {\bibinfo {volume} {95}},\
  \bibinfo {pages} {206603} (\bibinfo {year} {2005})}\BibitemShut {NoStop}%
\bibitem [{\citenamefont {Basko}\ \emph {et~al.}(2006)\citenamefont {Basko},
  \citenamefont {Aleiner},\ and\ \citenamefont {Altshuler}}]{Basko2006}%
  \BibitemOpen
  \bibfield  {author} {\bibinfo {author} {\bibfnamefont {D.}~\bibnamefont
  {Basko}}, \bibinfo {author} {\bibfnamefont {I.}~\bibnamefont {Aleiner}},\
  and\ \bibinfo {author} {\bibfnamefont {B.}~\bibnamefont {Altshuler}},\
  }\bibfield  {title} {\bibinfo {title} {Metal{\textendash}insulator transition
  in a weakly interacting many-electron system with localized single-particle
  states},\ }\href {https://doi.org/10.1016/j.aop.2005.11.014} {\bibfield
  {journal} {\bibinfo  {journal} {Annals of Physics}\ }\textbf {\bibinfo
  {volume} {321}},\ \bibinfo {pages} {1126} (\bibinfo {year}
  {2006})}\BibitemShut {NoStop}%
\bibitem [{\citenamefont {Basko}\ \emph {et~al.}(2007)\citenamefont {Basko},
  \citenamefont {Aleiner},\ and\ \citenamefont {Altshuler}}]{Basko2007}%
  \BibitemOpen
  \bibfield  {author} {\bibinfo {author} {\bibfnamefont {D.~M.}\ \bibnamefont
  {Basko}}, \bibinfo {author} {\bibfnamefont {I.~L.}\ \bibnamefont {Aleiner}},\
  and\ \bibinfo {author} {\bibfnamefont {B.~L.}\ \bibnamefont {Altshuler}},\
  }\bibfield  {title} {\bibinfo {title} {Possible experimental manifestations
  of the many-body localization},\ }\href
  {https://doi.org/10.1103/PhysRevB.76.052203} {\bibfield  {journal} {\bibinfo
  {journal} {Phys. Rev. B}\ }\textbf {\bibinfo {volume} {76}},\ \bibinfo
  {pages} {052203} (\bibinfo {year} {2007})}\BibitemShut {NoStop}%
\bibitem [{\citenamefont {Oganesyan}\ and\ \citenamefont
  {Huse}(2007)}]{Oganesyan2007}%
  \BibitemOpen
  \bibfield  {author} {\bibinfo {author} {\bibfnamefont {V.}~\bibnamefont
  {Oganesyan}}\ and\ \bibinfo {author} {\bibfnamefont {D.~A.}\ \bibnamefont
  {Huse}},\ }\bibfield  {title} {\bibinfo {title} {Localization of interacting
  fermions at high temperature},\ }\href
  {https://doi.org/10.1103/PhysRevB.75.155111} {\bibfield  {journal} {\bibinfo
  {journal} {Phys. Rev. B}\ }\textbf {\bibinfo {volume} {75}},\ \bibinfo
  {pages} {155111} (\bibinfo {year} {2007})}\BibitemShut {NoStop}%
\bibitem [{\citenamefont {Serbyn}\ \emph {et~al.}(2013)\citenamefont {Serbyn},
  \citenamefont {Papi\ifmmode~\acute{c}\else \'{c}\fi{}},\ and\ \citenamefont
  {Abanin}}]{serbyn2013local}%
  \BibitemOpen
  \bibfield  {author} {\bibinfo {author} {\bibfnamefont {M.}~\bibnamefont
  {Serbyn}}, \bibinfo {author} {\bibfnamefont {Z.}~\bibnamefont
  {Papi\ifmmode~\acute{c}\else \'{c}\fi{}}},\ and\ \bibinfo {author}
  {\bibfnamefont {D.~A.}\ \bibnamefont {Abanin}},\ }\bibfield  {title}
  {\bibinfo {title} {Local conservation laws and the structure of the many-body
  localized states},\ }\href {https://doi.org/10.1103/PhysRevLett.111.127201}
  {\bibfield  {journal} {\bibinfo  {journal} {Phys. Rev. Lett.}\ }\textbf
  {\bibinfo {volume} {111}},\ \bibinfo {pages} {127201} (\bibinfo {year}
  {2013})}\BibitemShut {NoStop}%
\bibitem [{\citenamefont {Huse}\ \emph {et~al.}(2014)\citenamefont {Huse},
  \citenamefont {Nandkishore},\ and\ \citenamefont
  {Oganesyan}}]{huse2014phenomenology}%
  \BibitemOpen
  \bibfield  {author} {\bibinfo {author} {\bibfnamefont {D.~A.}\ \bibnamefont
  {Huse}}, \bibinfo {author} {\bibfnamefont {R.}~\bibnamefont {Nandkishore}},\
  and\ \bibinfo {author} {\bibfnamefont {V.}~\bibnamefont {Oganesyan}},\
  }\bibfield  {title} {\bibinfo {title} {Phenomenology of fully
  many-body-localized systems},\ }\href
  {https://doi.org/10.1103/PhysRevB.90.174202} {\bibfield  {journal} {\bibinfo
  {journal} {Phys. Rev. B}\ }\textbf {\bibinfo {volume} {90}},\ \bibinfo
  {pages} {174202} (\bibinfo {year} {2014})}\BibitemShut {NoStop}%
\bibitem [{\citenamefont {Nandkishore}\ and\ \citenamefont
  {Huse}(2015)}]{nandkishore_review}%
  \BibitemOpen
  \bibfield  {author} {\bibinfo {author} {\bibfnamefont {R.}~\bibnamefont
  {Nandkishore}}\ and\ \bibinfo {author} {\bibfnamefont {D.~A.}\ \bibnamefont
  {Huse}},\ }\bibfield  {title} {\bibinfo {title} {Many-body localization and
  thermalization in quantum statistical mechanics},\ }\href
  {https://doi.org/10.1146/annurev-conmatphys-031214-014726} {\bibfield
  {journal} {\bibinfo  {journal} {Annual Review of Condensed Matter Physics}\
  }\textbf {\bibinfo {volume} {6}},\ \bibinfo {pages} {15} (\bibinfo {year}
  {2015})}\BibitemShut {NoStop}%
\bibitem [{\citenamefont {Abanin}\ \emph {et~al.}(2019)\citenamefont {Abanin},
  \citenamefont {Altman}, \citenamefont {Bloch},\ and\ \citenamefont
  {Serbyn}}]{Abanin2019}%
  \BibitemOpen
  \bibfield  {author} {\bibinfo {author} {\bibfnamefont {D.~A.}\ \bibnamefont
  {Abanin}}, \bibinfo {author} {\bibfnamefont {E.}~\bibnamefont {Altman}},
  \bibinfo {author} {\bibfnamefont {I.}~\bibnamefont {Bloch}},\ and\ \bibinfo
  {author} {\bibfnamefont {M.}~\bibnamefont {Serbyn}},\ }\bibfield  {title}
  {\bibinfo {title} {Colloquium: Many-body localization, thermalization, and
  entanglement},\ }\href {https://doi.org/10.1103/RevModPhys.91.021001}
  {\bibfield  {journal} {\bibinfo  {journal} {Rev. Mod. Phys.}\ }\textbf
  {\bibinfo {volume} {91}},\ \bibinfo {pages} {021001} (\bibinfo {year}
  {2019})}\BibitemShut {NoStop}%
\bibitem [{\citenamefont {Mirlin}(2000)}]{MIRLIN2000259}%
  \BibitemOpen
  \bibfield  {author} {\bibinfo {author} {\bibfnamefont {A.~D.}\ \bibnamefont
  {Mirlin}},\ }\bibfield  {title} {\bibinfo {title} {Statistics of energy
  levels and eigenfunctions in disordered systems},\ }\href
  {https://doi.org/https://doi.org/10.1016/S0370-1573(99)00091-5} {\bibfield
  {journal} {\bibinfo  {journal} {Physics Reports}\ }\textbf {\bibinfo {volume}
  {326}},\ \bibinfo {pages} {259} (\bibinfo {year} {2000})}\BibitemShut
  {NoStop}%
\bibitem [{\citenamefont {Serbyn}\ and\ \citenamefont
  {Moore}(2016)}]{Serbyn_2016}%
  \BibitemOpen
  \bibfield  {author} {\bibinfo {author} {\bibfnamefont {M.}~\bibnamefont
  {Serbyn}}\ and\ \bibinfo {author} {\bibfnamefont {J.~E.}\ \bibnamefont
  {Moore}},\ }\bibfield  {title} {\bibinfo {title} {Spectral statistics across
  the many-body localization transition},\ }\href
  {https://doi.org/10.1103/PhysRevB.93.041424} {\bibfield  {journal} {\bibinfo
  {journal} {Phys. Rev. B}\ }\textbf {\bibinfo {volume} {93}},\ \bibinfo
  {pages} {041424} (\bibinfo {year} {2016})}\BibitemShut {NoStop}%
\bibitem [{\citenamefont {Nandkishore}\ \emph {et~al.}(2014)\citenamefont
  {Nandkishore}, \citenamefont {Gopalakrishnan},\ and\ \citenamefont
  {Huse}}]{Huse_PRB_2014}%
  \BibitemOpen
  \bibfield  {author} {\bibinfo {author} {\bibfnamefont {R.}~\bibnamefont
  {Nandkishore}}, \bibinfo {author} {\bibfnamefont {S.}~\bibnamefont
  {Gopalakrishnan}},\ and\ \bibinfo {author} {\bibfnamefont {D.~A.}\
  \bibnamefont {Huse}},\ }\bibfield  {title} {\bibinfo {title} {Spectral
  features of a many-body-localized system weakly coupled to a bath},\ }\href
  {https://doi.org/10.1103/PhysRevB.90.064203} {\bibfield  {journal} {\bibinfo
  {journal} {Phys. Rev. B}\ }\textbf {\bibinfo {volume} {90}},\ \bibinfo
  {pages} {064203} (\bibinfo {year} {2014})}\BibitemShut {NoStop}%
\bibitem [{\citenamefont {Johri}\ \emph {et~al.}(2015)\citenamefont {Johri},
  \citenamefont {Nandkishore},\ and\ \citenamefont {Bhatt}}]{Jori_PRL_2015}%
  \BibitemOpen
  \bibfield  {author} {\bibinfo {author} {\bibfnamefont {S.}~\bibnamefont
  {Johri}}, \bibinfo {author} {\bibfnamefont {R.}~\bibnamefont {Nandkishore}},\
  and\ \bibinfo {author} {\bibfnamefont {R.~N.}\ \bibnamefont {Bhatt}},\
  }\bibfield  {title} {\bibinfo {title} {Many-body localization in imperfectly
  isolated quantum systems},\ }\href
  {https://doi.org/10.1103/PhysRevLett.114.117401} {\bibfield  {journal}
  {\bibinfo  {journal} {Phys. Rev. Lett.}\ }\textbf {\bibinfo {volume} {114}},\
  \bibinfo {pages} {117401} (\bibinfo {year} {2015})}\BibitemShut {NoStop}%
\bibitem [{\citenamefont {Manucharyan}\ \emph {et~al.}(2009)\citenamefont
  {Manucharyan}, \citenamefont {Koch}, \citenamefont {Glazman},\ and\
  \citenamefont {Devoret}}]{Manucharyan2009}%
  \BibitemOpen
  \bibfield  {author} {\bibinfo {author} {\bibfnamefont {V.~E.}\ \bibnamefont
  {Manucharyan}}, \bibinfo {author} {\bibfnamefont {J.}~\bibnamefont {Koch}},
  \bibinfo {author} {\bibfnamefont {L.~I.}\ \bibnamefont {Glazman}},\ and\
  \bibinfo {author} {\bibfnamefont {M.~H.}\ \bibnamefont {Devoret}},\
  }\bibfield  {title} {\bibinfo {title} {Fluxonium: Single {C}ooper-pair
  circuit free of charge offsets},\ }\href
  {https://doi.org/10.1126/science.1175552} {\bibfield  {journal} {\bibinfo
  {journal} {Science}\ }\textbf {\bibinfo {volume} {326}},\ \bibinfo {pages}
  {113} (\bibinfo {year} {2009})}\BibitemShut {NoStop}%
\bibitem [{\citenamefont {Meiser}\ and\ \citenamefont
  {Meystre}(2006)}]{Meystre_2006}%
  \BibitemOpen
  \bibfield  {author} {\bibinfo {author} {\bibfnamefont {D.}~\bibnamefont
  {Meiser}}\ and\ \bibinfo {author} {\bibfnamefont {P.}~\bibnamefont
  {Meystre}},\ }\bibfield  {title} {\bibinfo {title} {Superstrong coupling
  regime of cavity quantum electrodynamics},\ }\href
  {https://doi.org/10.1103/PhysRevA.74.065801} {\bibfield  {journal} {\bibinfo
  {journal} {Phys. Rev. A}\ }\textbf {\bibinfo {volume} {74}},\ \bibinfo
  {pages} {065801} (\bibinfo {year} {2006})}\BibitemShut {NoStop}%
\bibitem [{\citenamefont {Sundaresan}\ \emph {et~al.}(2015)\citenamefont
  {Sundaresan}, \citenamefont {Liu}, \citenamefont {Sadri}, \citenamefont
  {Sz\ifmmode~\mbox{\H{o}}\else \H{o}\fi{}cs}, \citenamefont {Underwood},
  \citenamefont {Malekakhlagh}, \citenamefont {T\"ureci},\ and\ \citenamefont
  {Houck}}]{Houck_2015}%
  \BibitemOpen
  \bibfield  {author} {\bibinfo {author} {\bibfnamefont {N.~M.}\ \bibnamefont
  {Sundaresan}}, \bibinfo {author} {\bibfnamefont {Y.}~\bibnamefont {Liu}},
  \bibinfo {author} {\bibfnamefont {D.}~\bibnamefont {Sadri}}, \bibinfo
  {author} {\bibfnamefont {L.~J.}\ \bibnamefont {Sz\ifmmode~\mbox{\H{o}}\else
  \H{o}\fi{}cs}}, \bibinfo {author} {\bibfnamefont {D.~L.}\ \bibnamefont
  {Underwood}}, \bibinfo {author} {\bibfnamefont {M.}~\bibnamefont
  {Malekakhlagh}}, \bibinfo {author} {\bibfnamefont {H.~E.}\ \bibnamefont
  {T\"ureci}},\ and\ \bibinfo {author} {\bibfnamefont {A.~A.}\ \bibnamefont
  {Houck}},\ }\bibfield  {title} {\bibinfo {title} {Beyond strong coupling in a
  multimode cavity},\ }\href {https://doi.org/10.1103/PhysRevX.5.021035}
  {\bibfield  {journal} {\bibinfo  {journal} {Phys. Rev. X}\ }\textbf {\bibinfo
  {volume} {5}},\ \bibinfo {pages} {021035} (\bibinfo {year}
  {2015})}\BibitemShut {NoStop}%
\bibitem [{\citenamefont {Mart{\'{\i}}nez}\ \emph {et~al.}(2019)\citenamefont
  {Mart{\'{\i}}nez}, \citenamefont {L{\'{e}}ger}, \citenamefont {Gheeraert},
  \citenamefont {Dassonneville}, \citenamefont {Planat}, \citenamefont
  {Foroughi}, \citenamefont {Krupko}, \citenamefont {Buisson}, \citenamefont
  {Naud}, \citenamefont {Hasch-Guichard}, \citenamefont {Florens},
  \citenamefont {Snyman},\ and\ \citenamefont {Roch}}]{PuertasMartnez2019}%
  \BibitemOpen
  \bibfield  {author} {\bibinfo {author} {\bibfnamefont {J.~P.}\ \bibnamefont
  {Mart{\'{\i}}nez}}, \bibinfo {author} {\bibfnamefont {S.}~\bibnamefont
  {L{\'{e}}ger}}, \bibinfo {author} {\bibfnamefont {N.}~\bibnamefont
  {Gheeraert}}, \bibinfo {author} {\bibfnamefont {R.}~\bibnamefont
  {Dassonneville}}, \bibinfo {author} {\bibfnamefont {L.}~\bibnamefont
  {Planat}}, \bibinfo {author} {\bibfnamefont {F.}~\bibnamefont {Foroughi}},
  \bibinfo {author} {\bibfnamefont {Y.}~\bibnamefont {Krupko}}, \bibinfo
  {author} {\bibfnamefont {O.}~\bibnamefont {Buisson}}, \bibinfo {author}
  {\bibfnamefont {C.}~\bibnamefont {Naud}}, \bibinfo {author} {\bibfnamefont
  {W.}~\bibnamefont {Hasch-Guichard}}, \bibinfo {author} {\bibfnamefont
  {S.}~\bibnamefont {Florens}}, \bibinfo {author} {\bibfnamefont
  {I.}~\bibnamefont {Snyman}},\ and\ \bibinfo {author} {\bibfnamefont
  {N.}~\bibnamefont {Roch}},\ }\bibfield  {title} {\bibinfo {title} {A tunable
  {J}osephson platform to explore many-body quantum optics in circuit-{QED}},\
  }\bibfield  {journal} {\bibinfo  {journal} {npj Quantum Information}\
  }\textbf {\bibinfo {volume} {5}},\ \href
  {https://doi.org/10.1038/s41534-018-0104-0} {10.1038/s41534-018-0104-0}
  (\bibinfo {year} {2019})\BibitemShut {NoStop}%
\bibitem [{\citenamefont {Kuzmin}\ \emph
  {et~al.}(2019{\natexlab{a}})\citenamefont {Kuzmin}, \citenamefont {Mehta},
  \citenamefont {Grabon}, \citenamefont {Mencia},\ and\ \citenamefont
  {Manucharyan}}]{Kuzmin2019}%
  \BibitemOpen
  \bibfield  {author} {\bibinfo {author} {\bibfnamefont {R.}~\bibnamefont
  {Kuzmin}}, \bibinfo {author} {\bibfnamefont {N.}~\bibnamefont {Mehta}},
  \bibinfo {author} {\bibfnamefont {N.}~\bibnamefont {Grabon}}, \bibinfo
  {author} {\bibfnamefont {R.}~\bibnamefont {Mencia}},\ and\ \bibinfo {author}
  {\bibfnamefont {V.~E.}\ \bibnamefont {Manucharyan}},\ }\bibfield  {title}
  {\bibinfo {title} {Superstrong coupling in circuit quantum electrodynamics},\
  }\bibfield  {journal} {\bibinfo  {journal} {npj Quantum Information}\
  }\textbf {\bibinfo {volume} {5}},\ \href
  {https://doi.org/10.1038/s41534-019-0134-2} {10.1038/s41534-019-0134-2}
  (\bibinfo {year} {2019}{\natexlab{a}})\BibitemShut {NoStop}%
\bibitem [{\citenamefont {Mehta}\ \emph {et~al.}(2022)\citenamefont {Mehta},
  \citenamefont {Ciuti}, \citenamefont {Kuzmin},\ and\ \citenamefont
  {Manucharyan}}]{Nitish_paper1}%
  \BibitemOpen
  \bibfield  {author} {\bibinfo {author} {\bibfnamefont {N.}~\bibnamefont
  {Mehta}}, \bibinfo {author} {\bibfnamefont {C.}~\bibnamefont {Ciuti}},
  \bibinfo {author} {\bibfnamefont {R.}~\bibnamefont {Kuzmin}},\ and\ \bibinfo
  {author} {\bibfnamefont {V.~E.}\ \bibnamefont {Manucharyan}},\ }\href@noop {}
  {\  (\bibinfo {year} {2022})}\BibitemShut {NoStop}%
\bibitem [{\citenamefont {Kuzmin}\ \emph
  {et~al.}(2019{\natexlab{b}})\citenamefont {Kuzmin}, \citenamefont {Mencia},
  \citenamefont {Grabon}, \citenamefont {Mehta}, \citenamefont {Lin},\ and\
  \citenamefont {Manucharyan}}]{Kuzmin2019_1}%
  \BibitemOpen
  \bibfield  {author} {\bibinfo {author} {\bibfnamefont {R.}~\bibnamefont
  {Kuzmin}}, \bibinfo {author} {\bibfnamefont {R.}~\bibnamefont {Mencia}},
  \bibinfo {author} {\bibfnamefont {N.}~\bibnamefont {Grabon}}, \bibinfo
  {author} {\bibfnamefont {N.}~\bibnamefont {Mehta}}, \bibinfo {author}
  {\bibfnamefont {Y.~H.}\ \bibnamefont {Lin}},\ and\ \bibinfo {author}
  {\bibfnamefont {V.~E.}\ \bibnamefont {Manucharyan}},\ }\bibfield  {title}
  {\bibinfo {title} {{Quantum electrodynamics of a superconductor–insulator
  phase transition}},\ }\href {https://doi.org/10.1038/s41567-019-0553-1}
  {\bibfield  {journal} {\bibinfo  {journal} {Nature Physics}\ }\textbf
  {\bibinfo {volume} {15}},\ \bibinfo {pages} {930} (\bibinfo {year}
  {2019}{\natexlab{b}})},\ \Eprint {https://arxiv.org/abs/1805.07379}
  {arXiv:1805.07379} \BibitemShut {NoStop}%
\bibitem [{\citenamefont {Nigg}\ \emph {et~al.}(2012)\citenamefont {Nigg},
  \citenamefont {Paik}, \citenamefont {Vlastakis}, \citenamefont {Kirchmair},
  \citenamefont {Shankar}, \citenamefont {Frunzio}, \citenamefont {Devoret},
  \citenamefont {Schoelkopf},\ and\ \citenamefont {Girvin}}]{Nigg_2012}%
  \BibitemOpen
  \bibfield  {author} {\bibinfo {author} {\bibfnamefont {S.~E.}\ \bibnamefont
  {Nigg}}, \bibinfo {author} {\bibfnamefont {H.}~\bibnamefont {Paik}}, \bibinfo
  {author} {\bibfnamefont {B.}~\bibnamefont {Vlastakis}}, \bibinfo {author}
  {\bibfnamefont {G.}~\bibnamefont {Kirchmair}}, \bibinfo {author}
  {\bibfnamefont {S.}~\bibnamefont {Shankar}}, \bibinfo {author} {\bibfnamefont
  {L.}~\bibnamefont {Frunzio}}, \bibinfo {author} {\bibfnamefont {M.~H.}\
  \bibnamefont {Devoret}}, \bibinfo {author} {\bibfnamefont {R.~J.}\
  \bibnamefont {Schoelkopf}},\ and\ \bibinfo {author} {\bibfnamefont {S.~M.}\
  \bibnamefont {Girvin}},\ }\bibfield  {title} {\bibinfo {title} {Black-box
  superconducting circuit quantization},\ }\href
  {https://doi.org/10.1103/PhysRevLett.108.240502} {\bibfield  {journal}
  {\bibinfo  {journal} {Phys. Rev. Lett.}\ }\textbf {\bibinfo {volume} {108}},\
  \bibinfo {pages} {240502} (\bibinfo {year} {2012})}\BibitemShut {NoStop}%
\bibitem [{\citenamefont {Kuzmin}\ \emph {et~al.}(2021)\citenamefont {Kuzmin},
  \citenamefont {Grabon}, \citenamefont {Mehta}, \citenamefont {Burshtein},
  \citenamefont {Goldstein}, \citenamefont {Houzet}, \citenamefont {Glazman},\
  and\ \citenamefont {Manucharyan}}]{Kuzmin_2021}%
  \BibitemOpen
  \bibfield  {author} {\bibinfo {author} {\bibfnamefont {R.}~\bibnamefont
  {Kuzmin}}, \bibinfo {author} {\bibfnamefont {N.}~\bibnamefont {Grabon}},
  \bibinfo {author} {\bibfnamefont {N.}~\bibnamefont {Mehta}}, \bibinfo
  {author} {\bibfnamefont {A.}~\bibnamefont {Burshtein}}, \bibinfo {author}
  {\bibfnamefont {M.}~\bibnamefont {Goldstein}}, \bibinfo {author}
  {\bibfnamefont {M.}~\bibnamefont {Houzet}}, \bibinfo {author} {\bibfnamefont
  {L.~I.}\ \bibnamefont {Glazman}},\ and\ \bibinfo {author} {\bibfnamefont
  {V.~E.}\ \bibnamefont {Manucharyan}},\ }\bibfield  {title} {\bibinfo {title}
  {Inelastic scattering of a photon by a quantum phase slip},\ }\href
  {https://doi.org/10.1103/PhysRevLett.126.197701} {\bibfield  {journal}
  {\bibinfo  {journal} {Phys. Rev. Lett.}\ }\textbf {\bibinfo {volume} {126}},\
  \bibinfo {pages} {197701} (\bibinfo {year} {2021})}\BibitemShut {NoStop}%
\bibitem [{\citenamefont {Gao}\ \emph {et~al.}(2019)\citenamefont {Gao},
  \citenamefont {Lester}, \citenamefont {Chou}, \citenamefont {Frunzio},
  \citenamefont {Devoret}, \citenamefont {Jiang}, \citenamefont {Girvin},\ and\
  \citenamefont {Schoelkopf}}]{gao2019entanglement}%
  \BibitemOpen
  \bibfield  {author} {\bibinfo {author} {\bibfnamefont {Y.~Y.}\ \bibnamefont
  {Gao}}, \bibinfo {author} {\bibfnamefont {B.~J.}\ \bibnamefont {Lester}},
  \bibinfo {author} {\bibfnamefont {K.~S.}\ \bibnamefont {Chou}}, \bibinfo
  {author} {\bibfnamefont {L.}~\bibnamefont {Frunzio}}, \bibinfo {author}
  {\bibfnamefont {M.~H.}\ \bibnamefont {Devoret}}, \bibinfo {author}
  {\bibfnamefont {L.}~\bibnamefont {Jiang}}, \bibinfo {author} {\bibfnamefont
  {S.}~\bibnamefont {Girvin}},\ and\ \bibinfo {author} {\bibfnamefont {R.~J.}\
  \bibnamefont {Schoelkopf}},\ }\bibfield  {title} {\bibinfo {title}
  {Entanglement of bosonic modes through an engineered exchange interaction},\
  }\href {https://www.nature.com/articles/s41586-019-0970-4} {\bibfield
  {journal} {\bibinfo  {journal} {Nature}\ }\textbf {\bibinfo {volume} {566}},\
  \bibinfo {pages} {509} (\bibinfo {year} {2019})}\BibitemShut {NoStop}%
\end{thebibliography}%


%apsrev4-2.bst 2019-01-14 (MD) hand-edited version of apsrev4-1.bst
%Control: key (0)
%Control: author (8) initials jnrlst
%Control: editor formatted (1) identically to author
%Control: production of article title (0) allowed
%Control: page (0) single
%Control: year (1) truncated
%Control: production of eprint (0) enabled
\begin{thebibliography}{3}%
\makeatletter
\providecommand \@ifxundefined [1]{%
 \@ifx{#1\undefined}
}%
\providecommand \@ifnum [1]{%
 \ifnum #1\expandafter \@firstoftwo
 \else \expandafter \@secondoftwo
 \fi
}%
\providecommand \@ifx [1]{%
 \ifx #1\expandafter \@firstoftwo
 \else \expandafter \@secondoftwo
 \fi
}%
\providecommand \natexlab [1]{#1}%
\providecommand \enquote  [1]{``#1''}%
\providecommand \bibnamefont  [1]{#1}%
\providecommand \bibfnamefont [1]{#1}%
\providecommand \citenamefont [1]{#1}%
\providecommand \href@noop [0]{\@secondoftwo}%
\providecommand \href [0]{\begingroup \@sanitize@url \@href}%
\providecommand \@href[1]{\@@startlink{#1}\@@href}%
\providecommand \@@href[1]{\endgroup#1\@@endlink}%
\providecommand \@sanitize@url [0]{\catcode `\\12\catcode `\$12\catcode
  `\&12\catcode `\#12\catcode `\^12\catcode `\_12\catcode `\%12\relax}%
\providecommand \@@startlink[1]{}%
\providecommand \@@endlink[0]{}%
\providecommand \url  [0]{\begingroup\@sanitize@url \@url }%
\providecommand \@url [1]{\endgroup\@href {#1}{\urlprefix }}%
\providecommand \urlprefix  [0]{URL }%
\providecommand \Eprint [0]{\href }%
\providecommand \doibase [0]{https://doi.org/}%
\providecommand \selectlanguage [0]{\@gobble}%
\providecommand \bibinfo  [0]{\@secondoftwo}%
\providecommand \bibfield  [0]{\@secondoftwo}%
\providecommand \translation [1]{[#1]}%
\providecommand \BibitemOpen [0]{}%
\providecommand \bibitemStop [0]{}%
\providecommand \bibitemNoStop [0]{.\EOS\space}%
\providecommand \EOS [0]{\spacefactor3000\relax}%
\providecommand \BibitemShut  [1]{\csname bibitem#1\endcsname}%
\let\auto@bib@innerbib\@empty
%</preamble>
\bibitem [{\citenamefont {Mehta}\ \emph {et~al.}(2022)\citenamefont {Mehta},
  \citenamefont {Ciuti}, \citenamefont {Kuzmin},\ and\ \citenamefont
  {Manucharyan}}]{Nitish_paper1}%
  \BibitemOpen
  \bibfield  {author} {\bibinfo {author} {\bibfnamefont {N.}~\bibnamefont
  {Mehta}}, \bibinfo {author} {\bibfnamefont {C.}~\bibnamefont {Ciuti}},
  \bibinfo {author} {\bibfnamefont {R.}~\bibnamefont {Kuzmin}},\ and\ \bibinfo
  {author} {\bibfnamefont {V.~E.}\ \bibnamefont {Manucharyan}},\ }\href@noop {}
  {\  (\bibinfo {year} {2022})}\BibitemShut {NoStop}%
\bibitem [{\citenamefont {Manucharyan}\ \emph {et~al.}(2009)\citenamefont
  {Manucharyan}, \citenamefont {Koch}, \citenamefont {Glazman},\ and\
  \citenamefont {Devoret}}]{Manucharyan2009}%
  \BibitemOpen
  \bibfield  {author} {\bibinfo {author} {\bibfnamefont {V.~E.}\ \bibnamefont
  {Manucharyan}}, \bibinfo {author} {\bibfnamefont {J.}~\bibnamefont {Koch}},
  \bibinfo {author} {\bibfnamefont {L.~I.}\ \bibnamefont {Glazman}},\ and\
  \bibinfo {author} {\bibfnamefont {M.~H.}\ \bibnamefont {Devoret}},\
  }\bibfield  {title} {\bibinfo {title} {Fluxonium: Single {C}ooper-pair
  circuit free of charge offsets},\ }\href
  {https://doi.org/10.1126/science.1175552} {\bibfield  {journal} {\bibinfo
  {journal} {Science}\ }\textbf {\bibinfo {volume} {326}},\ \bibinfo {pages}
  {113} (\bibinfo {year} {2009})}\BibitemShut {NoStop}%
\bibitem [{\citenamefont {Kuzmin}\ \emph {et~al.}(2019)\citenamefont {Kuzmin},
  \citenamefont {Mencia}, \citenamefont {Grabon}, \citenamefont {Mehta},
  \citenamefont {Lin},\ and\ \citenamefont {Manucharyan}}]{Kuzmin2019_1}%
  \BibitemOpen
  \bibfield  {author} {\bibinfo {author} {\bibfnamefont {R.}~\bibnamefont
  {Kuzmin}}, \bibinfo {author} {\bibfnamefont {R.}~\bibnamefont {Mencia}},
  \bibinfo {author} {\bibfnamefont {N.}~\bibnamefont {Grabon}}, \bibinfo
  {author} {\bibfnamefont {N.}~\bibnamefont {Mehta}}, \bibinfo {author}
  {\bibfnamefont {Y.~H.}\ \bibnamefont {Lin}},\ and\ \bibinfo {author}
  {\bibfnamefont {V.~E.}\ \bibnamefont {Manucharyan}},\ }\bibfield  {title}
  {\bibinfo {title} {{Quantum electrodynamics of a superconductor–insulator
  phase transition}},\ }\href {https://doi.org/10.1038/s41567-019-0553-1}
  {\bibfield  {journal} {\bibinfo  {journal} {Nature Physics}\ }\textbf
  {\bibinfo {volume} {15}},\ \bibinfo {pages} {930} (\bibinfo {year} {2019})},\
  \Eprint {https://arxiv.org/abs/1805.07379} {arXiv:1805.07379} \BibitemShut
  {NoStop}%
\end{thebibliography}%

\end{document}

\title{{\bf Supplementary Material for the article:}\\ ``Down-conversion of a single photon as a probe of many-body localization"}

\author{Nitish Mehta}
\affiliation{Department of Physics, University of Maryland, College Park, MD 20742, USA}

\author{Roman Kuzmin}
\affiliation{Department of Physics, University of Maryland, College Park, MD 20742, USA}

\author{ Cristiano Ciuti}
\affiliation{Laboratoire Mat\'{e}riaux et Ph\'{e}nom\`{e}nes Quantiques, Universit\'{e} Paris Cit\'{e}, CNRS-UMR7162, 75013 Paris, France}

\author{Vladimir E. Manucharyan}
\affiliation{Department of Physics, University of Maryland, College Park, MD 20742, USA}
\date{\today}
\maketitle
\usetagform{supplementary}
In this document we briefly summarize the theoretical model of our multi-mode quantum superconducting circuit as well as additional details on the data and experimental procedures.

\section{multi-mode
circuit QED}

The complete theoretical description of our device can be found in a separate theoretical work~\onlinecite{Nitish_paper1}.As it is common for light-matter coupling models, our circuit Hamiltonian can be separated into three terms:
\begin{equation}
    \hat{H}_{\textrm{QED}} = \hat{H}_{\mathrm{qubit}} + \hat{H}_{\mathrm{modes}} + \hat{H}_{\mathrm{int}}
    \label{H_full} \, .
    %\nonumber
\end{equation}

The first term $\hat{H}_{\mathrm{qubit}}$ defines the bare fluxonium circuit Hamiltonian as if it was disconnected from the transmission line:
\begin{equation}
    \hat{H}_{\mathrm{qubit}}/h = 4 E_C \hat{n}^2_J + \frac{1}{2} E_L \hat{\varphi}^2_J - E_J \cos (\hat{\varphi}_J - \varphi_{\mathrm{ext}}) \, .
    \label{H_fluxonium}
    %\nonumber
\end{equation}
The operators $\hat{n}_J$ and $\hat{\varphi}_J$ are the Cooper pair number and phase operators, respectively, describing the fluxonium circuit~\cite{Manucharyan2009}. They satisfy the commutation rule $[ \hat{\varphi}_J,\hat{n}_J]=\text{i}$. The parameters $E_C$ and $E_J$ are the charging and Josephson energies of the fluxonium's weak junction (labeled `1' in Fig.~1b of the main text), while the parameter $E_L$ is the inductive energy of the fluxonium's loop. In what follows, it is safe to truncate the Hamiltonian $\hat H_{\textrm{qubit}}$  to the lowest two energy levels which form the qubit transition at a flux-dependent frequency $f_{\mathrm{eg}} (\varphi_{\textrm{ext}})$. The second term $\hat{H}_{\mathrm{modes}}$ is used to introduce the discrete modes of the transmission line,
\begin{equation}
    \hat{H}_{\mathrm{modes}}/h = \sum_{i = 1}^{ +\infty} f_i^{(\mathrm{bare})} \hat{b}^{\dagger}_i \hat{b}_i  \, ,
    %\nonumber
\end{equation}
where the bosonic operators $\hat{b}_i, \hat{b}^{\dagger}_j$ obey the commutation rule $[\hat{b}_i,\hat{b}^{\dagger}_j] = \delta_{i,j}$. Finally, we chose the coupling term in the following form: 

\begin{equation}
    \hat{H}_{\mathrm{int}}/h = - \hat{\varphi}_J \sum_{i=1}^{j_0} g_i^{(f)} (\hat{b}_i + \hat{b}^{\dagger}_i) + \mathrm{i} \hat{n}_J \sum_{i=j_0+1}^{+\infty} g_i^{(c)} (\hat{b}^{\dagger}_i - \hat{b}_i ) \, ,
    %\nonumber
\end{equation}
that is, the coupling takes place via the flux variable for the low-frequency modes and via the charge coupling for the high-frequency modes. The mode index $j_0$ formally separates the low and the high frequencies ($j_0 = 20$ for the present device).\\

As further derived in Ref. \cite{Nitish_paper1}, the circuit Hamiltonian $\hat H_{\textrm{QED}}$ can be approximated by an effective Hamiltonian $\hat{H}$ (see Eqs. (1,2) of the main text, copied below for convenience):
 
\begin{equation}
    \frac{\hat{H}}{h} = \sum_{k>0} f_k \hat{a}_k^{\dagger} \hat{a}_k + \hat{V},
    \label{Heff1}
\end{equation}
\begin{equation}
    \hat{V} =g \sum_{j\leq j_0}^ {k,k'>j_0} \sqrt{j} \,A_{k,k'} \hat{a}_j^{\dagger} \hat{a}_{k'}^{\dagger} \hat{a}_k + h.c.\,
    \label{H_eff}
    %\nonumber
\end{equation}

This effective Hamiltonian describes a linear multi-mode cavity with flux-dependent standing-wave mode frequencies $f_k$ ($k = 1,2, ...$) and an additional three-wave mixing type non-linearity. Note that this effective model does not contain qubit variables,  but the parameters $f_k$ ($k = 1,2, ...$), $g$, and the matrix $A_{k,k'}$ do acquire a flux dependence, and they can be expressed in terms of a small number of circuit parameters summarized in Table 1.

\section{Single-particle spectrum}

Theoretically, the frequencies $f_k(\varphi_{\mathrm{ext}})$ define the single-particle excitation spectrum of our system. It is obtained by considering the linear hybridization of the qubit mode with the high-frequency modes (those with index $k > j_0$) of the transmission line, expressed via the following relation:
\begin{equation}
\hat{a}_k^{\dagger} \vert \mathrm{g} \rangle \vert 0 \rangle = 
W_{k,0} \vert \mathrm{e} \rangle \vert 0 \rangle + 
\sum_{k'} W_{k,k'} \vert \mathrm{g} \rangle  \hat{b}^{\dagger}_{k'} \vert 0 \rangle \, ,
%\nonumber
\end{equation} 
where the hybridization coefficients $W_{k,0}$ as well as the eigenfrequencies $f_k$ are determined by diagonalizing the circuit Hamiltonian $\hat{H}_{\textrm{QED}}$ in the basis of states
\begin{equation}
{\mathcal B }_0  = \left \{ \vert \mathrm{e} \rangle \vert 0 \rangle, \vert \mathrm{g} \rangle  \hat{b}^{\dagger}_{k} \vert 0 \rangle \right \}_{j_0 \leq k \leq N}  \, . 
%\nonumber
\end{equation}

It is assumed that only modes in the frequency window of width $\Gamma \ll f_{\mathrm{eg}}$ around the qubit frequency $f_{\mathrm{eg}} $ hybridize significantly. Consequently, the hybridization coefficient $W_{k, 0}$ goes to zero rapidly for $|f_k - f_{\mathrm{eg}}| \gtrsim \Gamma$.

\begin{figure*}[t!]
    \centering
    \includegraphics[width=1\linewidth]{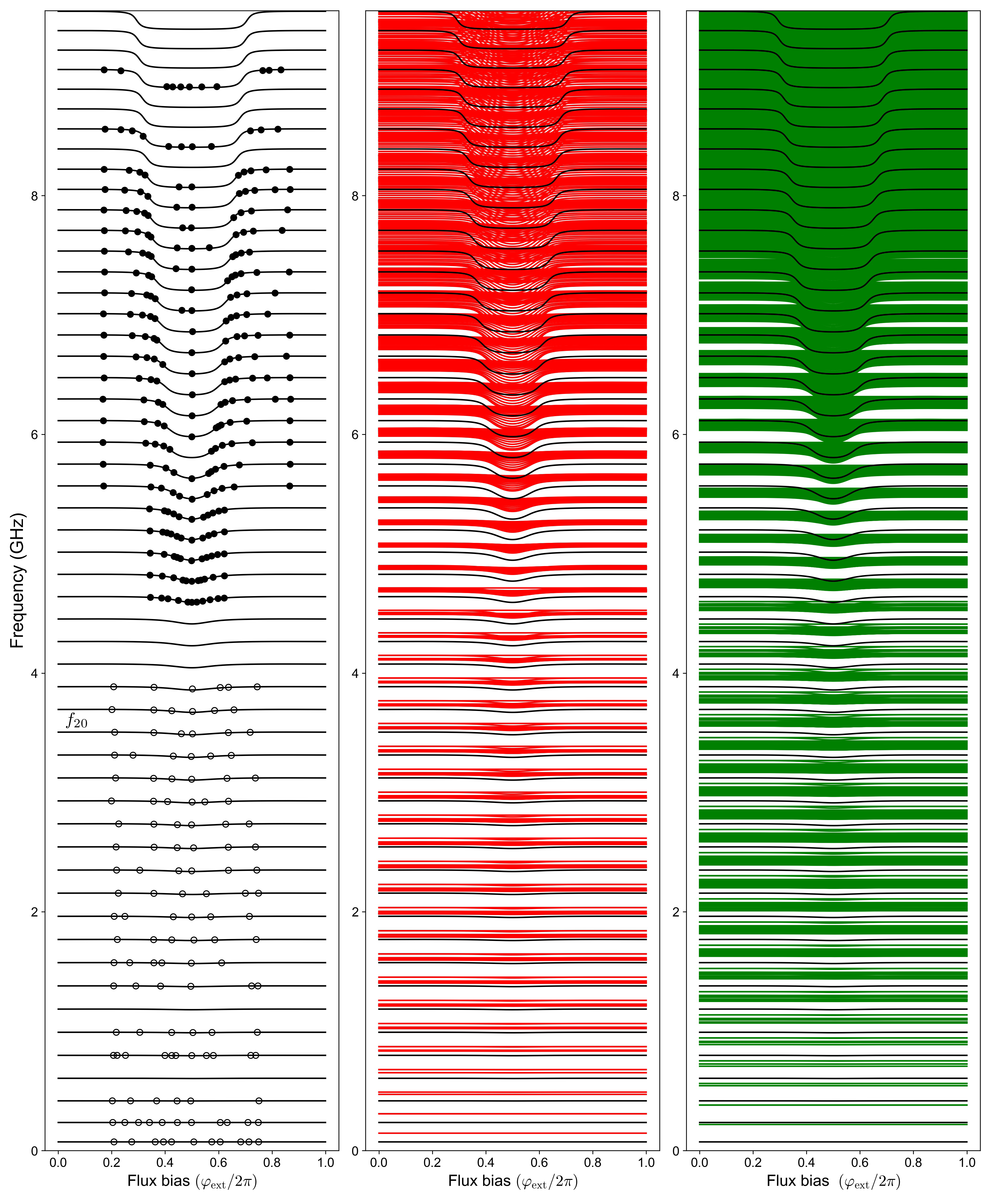}
    \caption{First column from the left:  excitation frequency versus the applied flux bias for the single-particle eigenstates (black solid lines). The filled black circles are the experimental data used for the fitting procedure described in the text. Second column: energy of the uncoupled two-particle states (red lines). Third column: same but for the  three-particle states (green lines).
    }
    \label{SM3}
\end{figure*}

\pagebreak
\newpage

This property justifies setting
$\hat{a}_k^{\dagger} = \hat{b}_k^{\dagger}$ for an appropriate cut-off $k \leq j_0$. Same considerations applies to the high-frequency cut-off $N$. The values  $j_0 = 20$ and $N = 200$ for the present device are chosen such that the standing wave mode frequencies $f_{k\leq j_0}$ and $f_{k > N}$ are insensitive to the tuning of the qubit frequency.

\section{Fitting $f_k$-data to circuit model}

Experimentally, the frequencies $f_k$ are the frequencies of the true standing wave modes of our coupled system. We directly measure these frequencies for $k = 1,2,...47$ using a combination of one-tone (at frequency above about $4~\textrm{GHz}$) and two-tone spectroscopy techniques, the procedure for which is described in Ref.~\cite{Kuzmin2019_1}. To extract circuit parameters from the measured single-particle spectra, we use the numerical expression for $f_k(\varphi_{\textrm{ext}})$, derived in Ref. \cite{Nitish_paper1}. Starting with the guess values for $E_J$, $E_C$, and $E_L$ as well as the measured values of the uncoupled standing wave mode frequencies $f^{(\mathrm{bare})}_k$ and transmission line parameters, 
we generate the theoretical values of  $  f_k(\varphi_{\mathrm{ext}})$. To reach good enough convergence and accuracy, we have used a cut-off mode number $N= 200$ and $j_0 = 20$. We then use a standard least mean squares minimization algorithm to find the optimal values for $(E_J, E_C, E_L)$ that fit the measured single-particle frequencies as a function of $\varphi_{\mathrm{ext}}$ (filled black markers in Fig. ~\ref{SM3}(a)). The markers are placed away from the two-particle splittings to get the best experimental estimate for the values of $f_k$. The accuracy of the resulting fit is close to the spectroscopic resolution limit. \\

Our procedure relies on the useful property $f^{(\mathrm{bare})}_k \approx f_k(\varphi_{\mathrm{ext}} = 0)$ in the frequency range of interest owing to the large qubit frequency detuning at the integer flux bias. Moreover, the data $f_k(\varphi_{\mathrm{ext}} = 0)$ allows us to extract the bare transmission line parameters: speed of light $v$, wave impedance $Z_{\infty}$, and the plasma propagation cut-off $f_p$, as explained in the next section, and hence only three parameters $E_J, E_C, E_L$ were used in the fit to spectroscopy data.   \\

With the full knowledge of the single-particle energies, we can readily generate the uncoupled multi-particle spectra Fig. ~\ref{SM3} (b) and (c). One must note two important properties of multi-particle spectra. First, even though there is no disorder in our system, the multi-particle states are spread out almost uniformly in energy at sufficiently large energy. This effect is due to the weak wave dispersion, discussed in the next section. Second, the multi-particle level spacing is much smaller than the single-particle level spacing already for two-particle states.

\section{Wave dispersion}
\begin{figure}[t!]
    \centering
    \includegraphics[width=\linewidth]{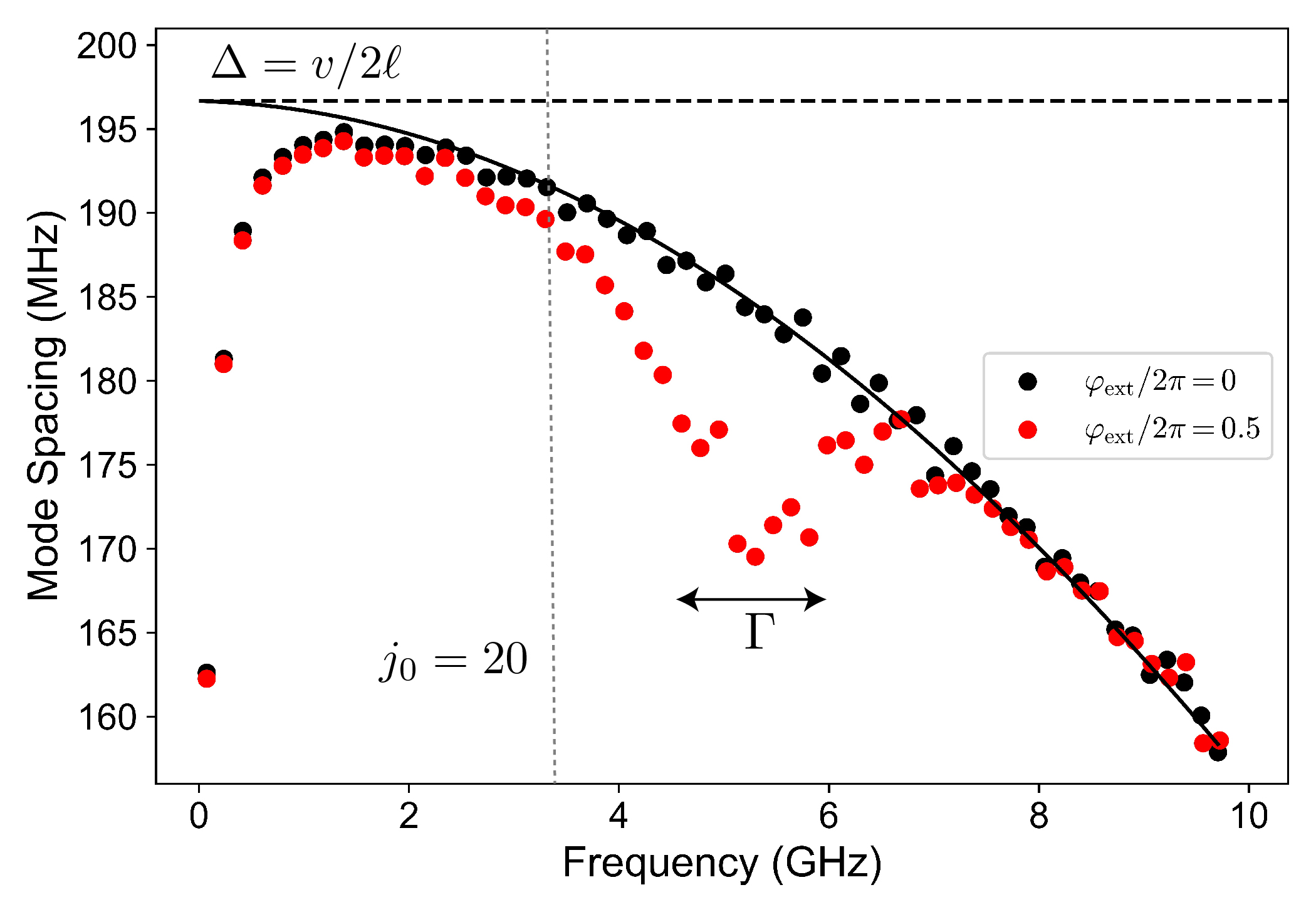}
    \caption{Measured standing wave mode frequency spacing $f_{k+1}-f_{k}$ versus frequency $f_k$ ($k = 1,2, ...$) at $\varphi_{\mathrm{ext}} = 0$ (black circles) and at $\varphi_{\mathrm{ext}}/2\pi = 0.5$ (red circles).
    The dispersion varies the mode spacing from $165$ to $195$ MHz in the frequency range of interest. The non-dispersive mode spacing is extracted to be $\Delta = v/2\ell \approx 197~\textrm{MHz}$. The hybridization window width $\Gamma \approx 1~\textrm{GHz}$ and the low-frequency cut-off corresponding to $j_0=20$ are indicated.
    }
    \label{FigSM1}
\end{figure}
We use the $f_k(\varphi_{\mathrm{ext}}=0)$ data to extract the wave propagation parameters of the transmission line. At $\varphi_{\mathrm{ext}}=0$ the qubit transition frequency $f_{\mathrm{eg}}$ is tuned above $15 \: \mathrm{GHz}$ and the qubit is effectively decoupled. This assumption is experimentally verified by the absence of flux-dependence of $f_k$ in the vicinity of $\varphi_{\mathrm{ext}}=0$ for the frequency range 0-10~\textrm{GHz}.

In such a decoupled configuration, the mode spacing obeys the known expression for the Josephson transmission line \cite{Kuzmin2019_1} (see Fig. ~\ref{FigSM1}):

\begin{equation}
   \Delta(f) =  \frac{v}{2 \ell} (1 - (f/f_p)^2)^{3/2} \, .
   \label{line_disp}
   %\nonumber
\end{equation}

Here  $\ell$ is the length of the transmission line and $f_p$ is the plasma frequency of the Josephson junctions, and $v$ is the speed of light in the long-wavelength limit. In the absence of dispersion, the mode spacing would be given by $\Delta = v/2\ell$. The plasma resonance prevents the propagation at frequencies above $f_p$, which is seen as the reduction in the mode spacing towards higher frequency. The mode spacing is also significantly distorted in the very low frequency limit due to frequency-dependent reflection phase at the antenna end. There is also a small irregularity in the frequency values, partly due to the inhomogeneity of the transmission line parameters and partly due to the small spectroscopic linewidth. This irregularity amounts to the frequency shifts of order $1\%$ of the mode spacing.\\  

It is also useful to directly compare the wave dispersion at $\varphi_{\textrm{ext}} = 0$ and $\varphi_{\textrm{ext}} = \pi$, where the multi-particle effects are also absent.  The dip in the mode spacing at about $5~\textrm{GHz}$ corresponds to the qubit frequency $f_{\textrm{eg}} (\varphi_{\textrm{ext}} = \pi) \approx 5~\textrm{GHz}$ and its width illustrates the hybridization window given by the parameter $\Gamma \approx 1~\textrm{GHz}$. Since we do most of the spectroscopy for the qubit frequency in the range $6-8~\textrm{GHz}$, the choice of the low-frequency cut-off $j_0 = 20$ or high-frequency cut-off $N=200$ is well justified.

\section{Photon-photon interaction}

\begin{figure}[t!]
    \centering
    \includegraphics[width=1\linewidth]{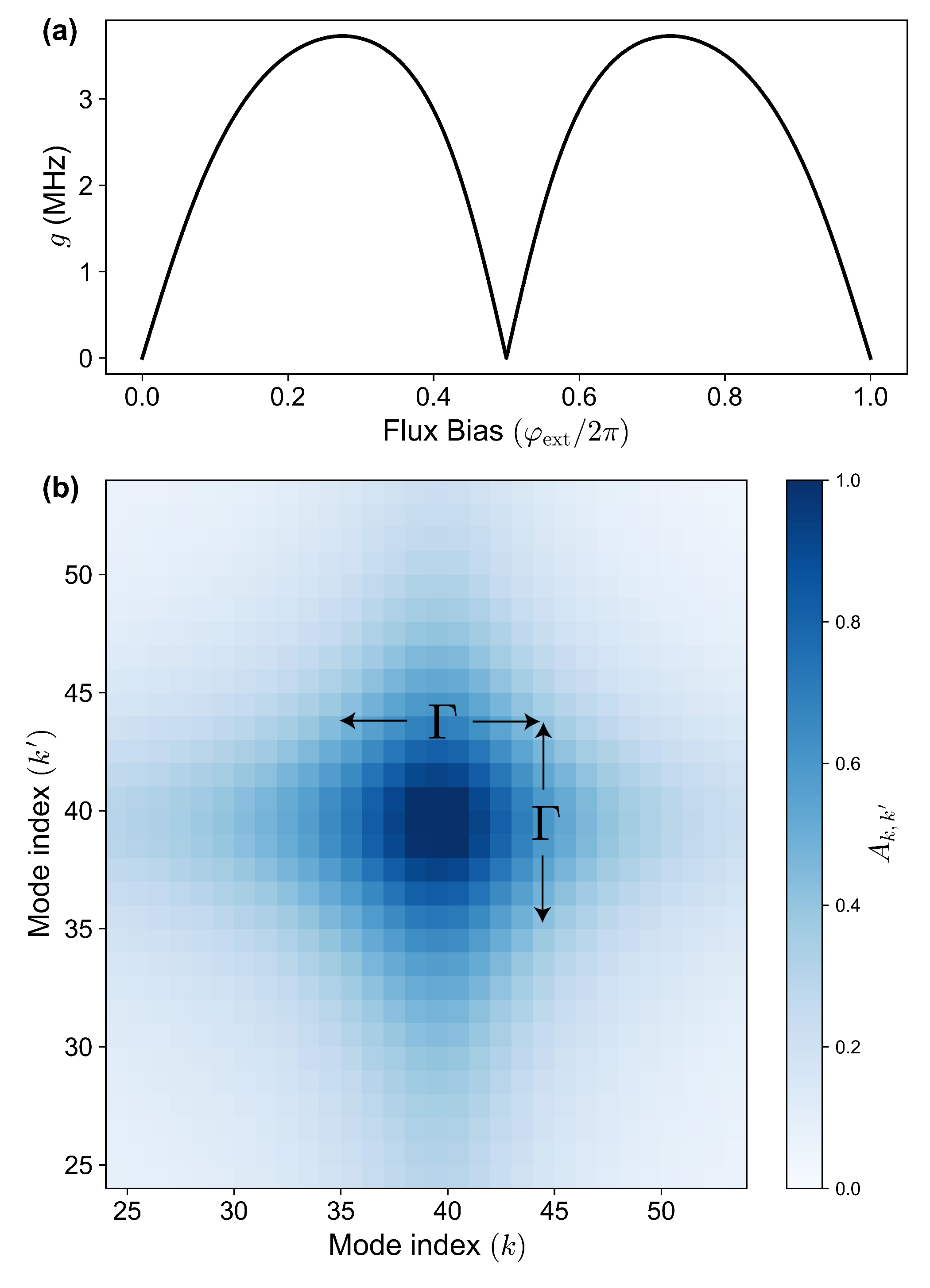}
    \caption{(a) Photon-photon interaction scale $g$ as a function of the external flux. (b) Color plot of the matrix elements $A_{k,k'}$ versus the mode indices $k, k'$ when the qubit frequency is tuned $f_{\mathrm{eg}} = 7.2 \: \mathrm{GHz}$ (around mode index 39) at $\varphi_{\mathrm{ext}}/ 2\pi = 0.36$. Note that the width of the local maximum in $A_{k,k'}$ is given by $\Gamma/\Delta$.
    }
    \label{FigSM2}
\end{figure}

The parameters of the interaction term $\hat{V}$ can be conveniently summarized by Fig. ~\ref{FigSM2}, which shows the overall energy scale $g (\varphi_{\textrm{ext}})$ (see Fig. ~ \ref{FigSM2}a) and the matrix $A_{k,k'}$ (see Fig. ~\ref{FigSM2}b), defined as follows:
\begin{equation}
    g \simeq \frac{\Delta^2}{\Gamma} \sqrt{\frac{h/(2e)^2}{32 \pi Z_{\infty} }} (\langle \mathrm{e} \vert \hat{\varphi}_J \vert \mathrm{e} \rangle -  \langle \mathrm{g} \vert \hat{\varphi}_J \vert \mathrm{g} \rangle) \, \label{g} ,
      %\nonumber
\end{equation}
\begin{equation}
    A_{k, k'} =  \frac{1}{\text{max}(W^{*}_{k,0} W_{k, 0})}W^{*}_{k,0} W_{k', 0}  \, . \label{Akk}
    %\nonumber
\end{equation}

The quantity $g$ sets the frequency scale for the largest possible two-particle splitting. The symmetry-breaking factor $\langle \mathrm{e} \vert \hat{\varphi}_J \vert \mathrm{e} \rangle -  \langle \mathrm{g} \vert \hat{\varphi}_J \vert \mathrm{g} \rangle\, \label{g}$ vanishes at $\varphi_{\text{ext}} = 0, \pi$ (see Fig. ~ \ref{FigSM2}a) and has a maximum value of approximately $3.5$. The expression for $A_{k, k'}$ is essentially the product of the linear qubit component in modes $k$ and $k'$. It is normalized in such a way that $A_{k,k} = 1$ for $f_k = f_{\mathrm{eg}}$. The quantity $\text{max}(W^{*}_{k,0} W_{k, 0}) \sim \Delta/\Gamma$, and it is achieved by tuning the qubit frequency such that $f_{\mathrm{eg}} \approx f_k \pm \Gamma$.

To determine the final spectral structure reported in the figures of the main text, we have diagonalized the effective Hamiltonian in an enlarged Hilbert space containing multi-particle states. We stress that this has been obtained with the same parameters extracted from the dispersion of the single-particle states. If we wish to calculate the spectral function around mode $k$ including two-particle states, we need to diagonalize the effective Hamiltonian in the following basis:
\begin{equation}
 \left \{ 
\hat{a}^{\dagger}_{k} 
\vert 0 \rangle \right \} \bigcup 
\left \{ 
\hat{a}^{\dagger}_{k'} \hat{a}^{\dagger}_{i} 
\vert 0 \rangle \right \}^{j_{0} < k' < k}_{1 \leq i \leq j_{0}}   \, .
   \nonumber
\end{equation}

For calculations including also three-particle states, the basis is enlarged as follows:
\begin{equation}
 \left \{ 
\hat{a}^{\dagger}_{k} 
\vert 0 \rangle \right \} \bigcup 
\left \{ 
\hat{a}^{\dagger}_{k'} \hat{a}^{\dagger}_{i} 
\vert 0 \rangle \right \}^{j_{0} < k' <  k}_{1 \leq i \leq j_{0}} 
\bigcup 
\left \{ 
\hat{a}^{\dagger}_{k'} \hat{a}^{\dagger}_{k''} \hat{a}^{\dagger}_{i} 
\vert 0 \rangle \right \}^{j_{0} < k',k'' < k}_{1 \leq i \leq j_{0}}
\, .
   \nonumber
\end{equation}
The typical size of the matrices to get good convergence in the plots reported in the main text is of the order $100 \times 100$. This procedure can be iterated to include higher-order multi-particle states.

\section{Calculation of $|S_{11}|$}

\begin{figure}[t!]
    \centering
    \includegraphics[width=1\linewidth]{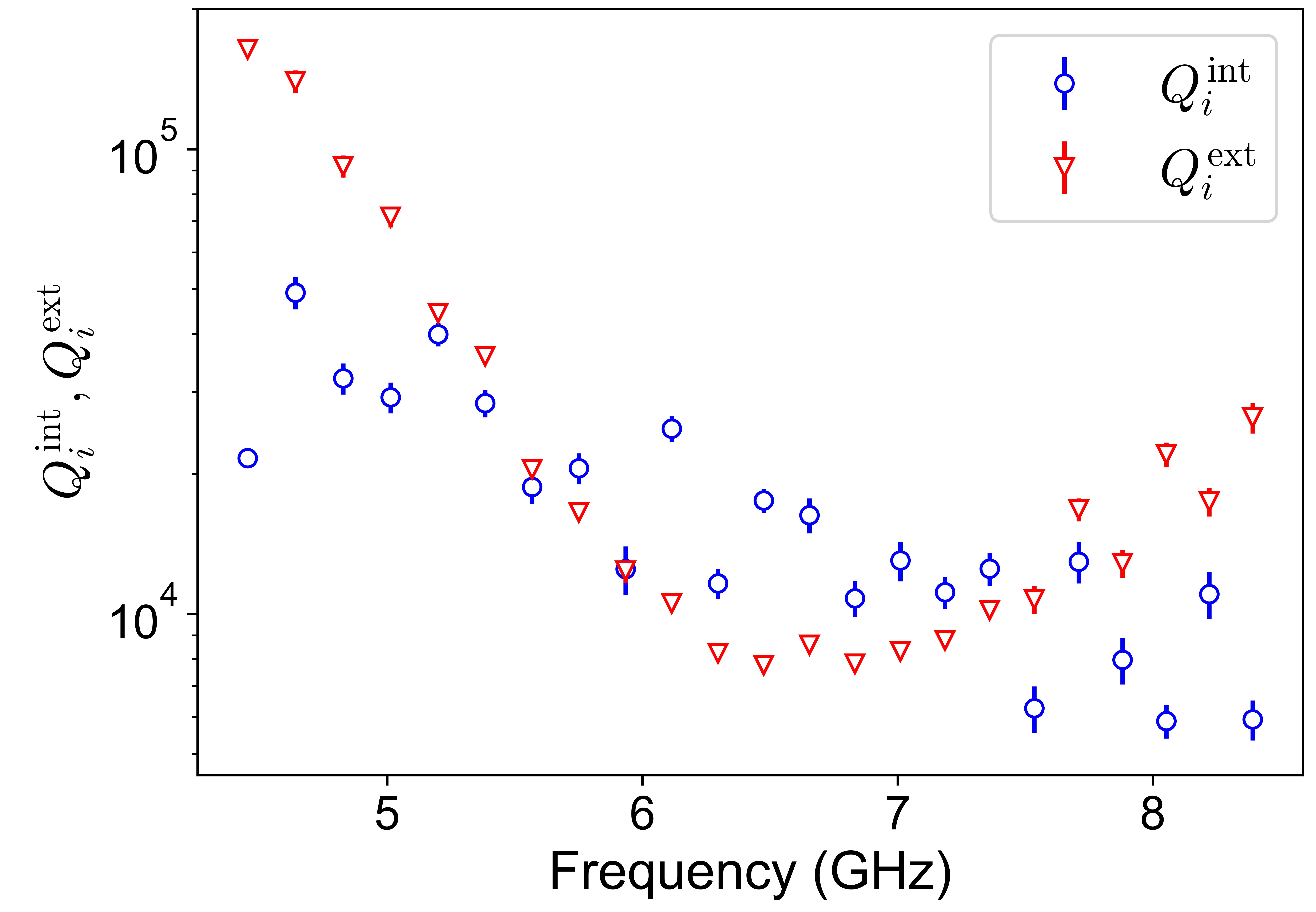}
    \caption{Internal (blue circles) and external (red triangles) quality factors of the bare Fabry-Perot modes as a function of mode frequency measured at the  flux bias $\varphi_{\mathrm{ext}}= 0$. For this flux value, the qubit coupling is negligible. 
    }
    \label{FigSM5}
\end{figure}

In order to theoretically calculate the linear microwave reflection spectra, we first consider the expression for the bare transmission line:
%\begin{widetext}
\begin{equation}
S_{11} (f) =\prod_{i=1}^{N} \frac{2 \mathrm{i} \Big(\frac{f - f^{(\mathrm{bare})}_i}{ f^{(\mathrm{bare})}_i} \Big)- \frac{1}{Q^{{\mathrm{ext}}}_{i}} + \frac{1}{Q^{{\mathrm{int}}}_{i}} } {2 \mathrm{i} \Big(\frac{f - f^{(\mathrm{bare})}_i}{ f^{(\mathrm{bare})}_i}\Big) + \frac{1}{Q^{{\mathrm{ext}}}_{i}} + \frac{1}{Q^{{\mathrm{int}}}_{i}}}\, ,
\label{ref_coe_bare}
%\nonumber
\end{equation} 
 where $Q^{\mathrm{int}}_{i}$ and $Q^{\mathrm{ext}}_{i}$ are respectively the internal and external quality factors of the bare transmission line modes (see Fig. ~\ref{FigSM5}). To calculate the reflection coefficient in the presence of the qubit, the expression is replaced by the product over the many-body eigenstates of the effective Hamiltonian in Eq. (1) of the main text: 
\begin{equation}
S_{11} (f) =\prod_{i=1}^{N^{(s)}} \frac{2 \mathrm{i} \Big(\frac{f - f^{(many-body)}_i}{ f^{(many-body)}_i} \Big)- \frac{1}{\tilde{Q}^{{\mathrm{ext}}}_{i}}\ + \frac{1}{\tilde{Q}^{{\mathrm{int}}}_{i}} } {2 \mathrm{i} \Big(\frac{f - f^{(many-body)}_i}{ f^{(many-body)}_i} \Big) + \frac{1}{\tilde{Q}^{{\mathrm{ext}}}_{i}} + \frac{1}{\tilde{Q}^{{\mathrm{int}}}_{i}}}
\label{ref_coe},
%\nonumber
\end{equation} 
where ${f}^{(many-body)}_i$ are the excitation frequencies (with respect to the ground state) with corresponding eigenfunctions $\vert {\Psi}_i \rangle$ and $N^{(s)}$ are the total number of eigenstates of the Hamiltonian in the $s$-particle Hilbert subspace.
%\end{widetext}
Note that the external $(\tilde{Q}^{\mathrm{ext}}_{i})$ and internal $(\tilde{Q}^{\mathrm{int}}_{i})$  quality factors of the the many-body states depend on the single-particle weight in the many-body eigenstate. This can be calculated by weighing the eigenstate quality factors with their single-particle fractions, namely:

\begin{equation}
    (\tilde{Q}^{\mathrm{int} (\mathrm{ext})}_{i})^{-1} = \sum_{k> i_{0}} |\langle {\Psi}_i \vert \hat{a}^{\dagger}_{k} \vert 0 \rangle|^2 ({Q'}^{\mathrm{int} (\mathrm{ext})}_{k})^{-1}\,.
    %\nonumber
\end{equation}
Here the single-particle quality factors depend on the bare transmission line mode quality factors given by:
\begin{equation}
   ({Q'}^{\mathrm{int} (\mathrm{ext})}_{k})^{-1} = \sum_{i > i_{0}}   |W_{k ,i}|^2 (Q^{\mathrm{int} (\mathrm{ext})}_{i})^{-1} \, .
   %\nonumber
\end{equation}

\section{Summary of device parameters}

\begin{center}
\begin{table}[h!]
\begin{tabular}{||l | r||}
 \hline
 Parameter & Fitted value  \\ [0ex] 
 \hline\hline
 Speed of light $v$ & $2.36 \pm 0.01 \times 10^{6} \: \mathrm{m/s}$ \\ 
 \hline
Wave impedance $Z_{\infty}$ & $8.97 \pm 0.67 \: \mathrm{k \Omega}$  \\
 \hline
 Plasma frequency $f_p$ & $26.94  \pm 0.17\: \mathrm{GHz}$  \\
 \hline
 Charging energy $E_C$  & $8.0 \pm 0.4 \: \mathrm{GHz}$ \\ [1ex] 
 \hline
 Josephson energy $E_J$  & $8.9 \pm 0.6 \: \mathrm{GHz}$ \\ [1ex] 
 \hline
  Inductive energy $E_L$  & $1.39 \pm 0.05 \: \mathrm{GHz}$ \\ [1ex] 
 \hline
\end{tabular}
\caption{Device parameters obtained from fitting the $f_k(\varphi_{\textrm{ext}})$ theory to data. Additional parameters exactly defined by the fabrication are the transmission line length ($\ell = 6 \: \mathrm{mm}$) and the fraction of the fluxonium loop inductance shared with the transmission line (1/2).}
\end{table}
\end{center}

\newpage

\bibliography{biblio}